\documentclass[a4papers,12pt]{article}

\usepackage{amsmath,amssymb}   
\usepackage{epsfig}                    
\usepackage[matrix,arrow]{xy}          
\usepackage{xspace}                    
\usepackage{stmaryrd}                  
\usepackage{slashed}

\usepackage{cite}

\usepackage{hyperref} 

\usepackage{color,graphicx}

\numberwithin{equation}{section}       

\allowdisplaybreaks                    

\newcommand{\half}{{{\textstyle\frac{1}{2}}}}

\newcommand{\be}{\setlength\arraycolsep{2pt}\begin{equation}}
\newcommand{\ee}{\end{equation} }
\newcommand{\ba}{\begin{array}}
\newcommand{\ea}{\end{array}}
\newcommand{\bal}{\begin{aligned}}
\newcommand{\eal}{\end{aligned}}
\newcommand{\bpm}{\begin{pmatrix}}
\newcommand{\epm}{\end{pmatrix}}

\newcommand{\comm}[2]{\left[#1,#2\right]}
\newcommand{\order}[1]{\scriptscriptstyle (#1)}

\newcommand\alphap{{\alpha^{\prime}}}

\newcommand\tr{{\rm tr}}
\newcommand\Tr{{\rm Tr}}

\newcommand\cA{{\cal A}}

\newcommand\cD{{\cal D}}

\newcommand\cF{{\cal F}}

\newcommand\cH{{\cal H}}

\newcommand\cJ{{\cal J}}
\newcommand\cK{{\cal K}}
\newcommand\cL{{\cal L}}

\newcommand\cP{{\cal P}}

\newcommand\cS{{\cal S}}

\newcommand\cV{{\cal V}}

\newcommand\bcP{{\bar{\cP}}}

\newcommand\hcL{{\hat{\cal L}}}
\newcommand{\hcLo}{\hat{\cal L}^{\scriptscriptstyle 0}}

\newcommand\hn{\hat{n}}

\newcommand\hM{\hat{M}}
\newcommand\hN{\hat{N}}
\newcommand\hP{\hat{P}}
\newcommand\hQ{\hat{Q}}
\newcommand\hR{\hat{R}}
\newcommand\hS{\hat{S}}

\newcommand\hX{\hat{X}}

\newcommand\hcD{\hat{\cD}}

\def\brvare{\bar{\varepsilon}}
\def\breta{\bar{\eta}}

\def\brrho{\bar{\rho}}
\def\brchi{\bar{\chi}}
\def\brpsi{\bar{\psi}}

\def\bromega{{\bar{\omega}}}

\def\brPhi{{{\bar{\Phi}}}}

\def\bra{\bar{a}}
\def\brb{\bar{b}}
\def\brc{\bar{c}}
\def\bre{\bar{e}}

\def\brm{{\bar{m}}}
\def\brn{{\bar{n}}}
\def\brp{{\bar{p}}}
\def\brq{{\bar{q}}}
\def\brr{{\bar{r}}}
\def\brs{{\bar{s}}}

\def\brP{{\bar{P}}}
\def\brR{\bar{R}}

\def\brV{{\bar{V}}}

\def\hbrm{\hat{\bar{m}}}
\def\hbrn{\hat{\bar{n}}}
\def\hbrp{\hat{\bar{p}}}
\def\hbrq{\hat{\bar{q}}}

\def\hbrV{\hat{\bar{V}}}

\newcommand{\SO}{\mathbf{SO}}
\newcommand{\Spin}{\mathbf{Spin}}

\newcommand{\ODD}{\mathbf{O}(D,D)}
\newcommand{\ODDG}{\mathbf{O}(D,D+{\rm dim}\,G)}
\newcommand{\Odd}{\mathbf{O}(d,d)}

\newcommand{\ODG}{{\mathbf{O}(D{-1},1 + \text{dim}\, G)}}

\newcommand{\OoD}{{\mathbf{O}(1,D{-1})}}

\newcommand{\Gammao}{\Gamma^{\scriptscriptstyle{0}}{}}

\begin{document}
\begin{titlepage}
\begin{flushright}
Q15001
\end{flushright}

\vfill

\begin{center}
   \baselineskip=16pt
   {\LARGE \bf  Quadratic $\alphap$-Corrections to
   \\  Heterotic Double Field Theory 
   \\~\\~}
\\~\\
 \bf Kanghoon Lee \footnote{\tt kanghoon@kias.re.kr}
       \vskip .6cm
             \begin{small}
             	\vspace{2mm}
		{\it Korea Institute for Advanced Study, Seoul 130-722, Korea. }\\
\vspace{2mm}
\end{small}
\end{center}

\vfill 
\begin{center} 
\textbf{Abstract}
\end{center} 
\begin{quote}

We investigate $\alpha'$-corrections of heterotic double field theory up to quadratic order in the language of supersymmetric $\ODDG$ gauged double field theory. After introducing double-vielbein formalism with a parametrization which reproduces heterotic supergravity, we show that supersymmetry for heterotic double field theory up to leading order $\alpha'$-correction is obtained from supersymmetric gauged double field theory. We discuss the necessary modifications of the symmetries defined in supersymmetric gauged double field theory. Further, we construct supersymmetric completion at quadratic order in $\alpha'$. 
\end{quote} 
\vfill
\setcounter{footnote}{0}
\end{titlepage}
\newpage
\tableofcontents 

\section{Introduction}
Stringy correction in the low-energy effective string theories is described by supergravity with infinite number of higher-derivative terms. The higher-derivative structure is not arbitrary, but strongly restricted by supersymmetry of string theory.  For instance in heterotic supergravity the higher derivative corrections have been constructed up to cubic order in $\alphap$ through the supersymmetric completion \cite{Bergshoeff:1988nn,Bergshoeff:1989de,deRoo:1992zp}, and it has been shown that these results are consistent with the string amplitude calculations (see \cite{Peeters:2000qj,Tseytlin:1995bi} and references therein). However, the supersymmetric completion is an extremely difficult task due to the complexity of the structure. 

Another important ingredient for constructing the higher derivative correction is sting/$M$-theory duality. $\ODD$ $T$-duality or $\mathbf{SL}(2,\mathbb{Z})$ $S$-duality also provides strong constraint to string effective theory. Double field theory (DFT) with a section condition provides a manifest $\ODD$ T-duality covariant reformulation of the string low energy effective theory \cite{Siegel:1993xq,Siegel:1993th,Hull:2009mi,Hull:2009zb,Hohm:2010jy,Hohm:2010pp}. 
It has been applied to describe heterotic supergravity \cite{Hohm:2011ex}, type II supergravity \cite{Jeon:2012kd,Jeon:2012hp,Hohm:2011zr,Hohm:2011dv,Coimbra:2011nw} by incorporating Ramond-Ramond sector, M-thoery \cite{Berman:2010is,Berman:2011pe,Berman:2011cg,Berman:2011jh,Coimbra:2011ky,Coimbra:2012af} and gauged DFT\cite{Geissbuhler:2011mx,Aldazabal:2011nj,Grana:2012rr,Berman:2012uy,Musaev:2013rq,Dibitetto:2012rk,Berman:2013cli} which corresponds to lower dimensional gauged supergravity \cite{hermann1,hermann2}. Based on the geometric formulations involving local frame field \cite{Jeon:2010rw,Jeon:2011cn,Hohm:2010xe,Jeon:2011vx}, supersymmetric DFT has been constructed \cite{Hohm:2011nu,Jeon:2011sq}. 
However, there are obstructions when we try to construct $\alphap$-correction in the ordinary DFT. The main difficulty is the absence of rank-4 generalized tensor containing the Riemann tensor \cite{Jeon:2010rw,Jeon:2011cn,Coimbra:2011nw}. Moreover, it has been shown that $R^2$ term is forbidden in terms of the generalized metric \cite{Hohm:2011si}. 

Recent works have addressed the construction of $\alphap$-corrections for heterotic DFT in terms of $\ODDG$ gauged DFT by Bedoya, Marques and Nunez \cite{Bedoya:2014pma}.\footnote{ Another approach for $\alphap$-correction in DFT has been proposed by Hohm, Siegel and Zwiebach \cite{Hohm:2013jaa,Hohm:2014xsa}. In their approach,  generalized Lie derivative receives an $\alpha'$-correction instead of $\ODD$ T-duality transformation. However, in our paper we will focus on the gauged DFT approach.}
Similar approach in generalized geometry for stringy corrections has been studied by Coimbra, Minasian, Triendl and Waldram \cite{Coimbra:2014qaa}. The main idea is that $\SO (32)$ or $E_8\times E_8$ heterotic gauge group is enhanced by including the $\Spin(9,1)$ local Lorentz group, and their gauge fields are treated on an equal footing \cite{Andriot:2011iw,Garcia-Fernandez:2013gja,Baraglia:2013wua,Anderson:2014xha,delaOssa:2014cia,delaOssa:2014msa}. Since the gauge field for local Lorentz transformation is just spin-connection, the $R^2$ term naturally arises from the gauge kinetic term. They have also shown that the anomaly cancelation condition is given by Bianchi identity of the generalized curvature tensor. 

In the present paper, utilizing the double-vielbein formalism for the supersymmetric gauged DFT \cite{Berman:2013cli}, we investigate the supersymmetric structure of heterotic DFT up to quadratic order in $\alpha'$, and we examine the validity of gauged DFT description in higher order corrections. 
If we neglect $\alpha'$-corrections, heterotic DFT is identical with gauged DFT regardless of double-vielbein parametrization.
However in order to describe $\alpha'$-corrections, we should require a suitable parametrization which identifies the gauge field for $\mathbf{O}(D-1,1)$ local Lorentz group with the DFT spin-connection.
Therefore, in this paper heterotic DFT implies $\ODDG$ gauged DFT with a double-vielbein parametrization.
From defining properties of double-vielbein, we construct a consistent parametrization which provides a consistent description of heterotic supergravity. However, it is important to note that symmetries defined in gauged DFT do not preserve the parametrization.

For twisted generalized Lie derivative (\ref{generalizedLieDeriv}) in gauged DFT, we should lock the twisted generalized Lie derivative with the $\mathbf{O}(G)$ subgroup of $\ODG$ local Lorentz transformation in order to sustain the parametrization \cite{Bedoya:2014pma}. Then the twisted generalized Lie derivative is modified by the compensating $\mathbf{O}(G)$  local Lorentz transformation and the $\mathbf{O}(G)$ symmetry is broken.
For $\ODDG$ duality transformation, it is known that the global duality symmetry is broken to $\ODD$ subgroup due to the parametrization of $\ODDG$-covariant structure constant for the enhanced heterotic gauge group $G$ \cite{Hohm:2011ex, Hohm:2014sxa}. In addition, to preserve the parametrization of double-vielbein, the remaining global symmetry should also be modified by incorporating a compensating local Lorentz transformation.
As twisted generalized Lie derivative, we show that SUSY transformation for the leading order $\alphap$-correction is locked with the off-diagonal part of the $\ODG$ local Lorentz transformation. 

We also investigate supersymmetry for heterotic DFT linear in $\alpha'$. As the other bosonic symmetries, the supersymmetry transformation from gauged DFT should be modified to preserve the parametrization of double-vielbein. In \cite{Coimbra:2014qaa}, supersymmetry for heterotic supergravity has been constructed in the context of generalized geometry with the corrections linear in $\alphap$, and we discuss the relation with our result.

Another main result in this paper is the construction of supersymmetric completion at the quadratic order of $\alphap$-correction. As pointed out in \cite{Bergshoeff:1989de, Coimbra:2014qaa}, there exists a hidden higher order of $\alphap$-correction in the supersymmetry transformation of gravitino-curvature $\psi_{\brm\brn}$, and it leads $(\alphap)^2$-order corrections in the SUSY variation of the action. 
We then construct corrections of the action and SUSY transformation in order for canceling the $(\alphap)^2$ terms which arise from the SUSY variation of the gravitino-curvature.
Also, we show that there is no $(\alphap)^2$-correction including the cubic order of Riemann tensor, and it is in agreement with the earlier heterotic supergravity result \cite{Bergshoeff:1989de}.

The organization of the present paper is as follows. In section 2, we review heterotic DFT with $\alpha'$-correction \cite{Bedoya:2014pma} and double-vielbein formalism for $\ODDG$ gauged DFT \cite{Berman:2013cli} with an explicit parametrization. We show that the bosonic symmetries defined in the gauged DFT are modified to be consistent with the parametrization. Furthermore, connections and curvature tensors are introduced from gauged DFT. 
In section 3, fermion degrees of freedom are introduced and supersymmetry is constructed at linear order in $\alphap$. We show that SUSY transformation defined in $\ODDG$ gauged DFT is also modified by the parametrization. We end in section 4 by constructing supersymmetric completion at order $(\alphap)^2$.

\section{Double-vielbein formalism for heterotic DFT}
In the construction of $\alphap$-correction of the heterotic DFT \cite{Bedoya:2014pma, Coimbra:2014qaa}, one starts from $\ODDG$ gauged DFT which is the gauge group $G$ specified as
\be
G = G_1 \times G_2\,,
\ee
where $G_{1}$ is the $\SO(32)$ or $E_8\times E_8$ group for the heterotic Yang-Mills gauge symmetry and $G_{2}$ is the $\SO (9,1)$ local Lorentz group which acts on adjoint representation.
This results in a theory in which the heterotic gauge group and $\mathbf{O}(9,1)$ symmetry are treated on an equal footing \cite{Bergshoeff:1989de}.

In this section, we review heterotic DFT with $\alpha'$-corrections and double-vielbein formalism for $\ODDG$ gauged DFT with a suitable parametrization. 
We discuss how bosonic symmetries defined in gauged DFT are modified under the parametrization for double-vielbein. 
In addition we introduce the geometric quantities in gauged DFT.
There are several approaches for the geometric structure of gauged DFT \cite{Geissbuhler:2013uka, bermannew, Berman:2013cli}. Here we follow the so called semi-covariant approach \cite{Berman:2013cli} which is well-suited for supersymmetry.

\subsection{Double vielbein}
Suppose that heterotic DFT is defined on a generalized parallelizable space \cite{Lee:2014mla} to avoid a topological obstruction.
From the double-vielbein formalism of gauged DFT \cite{Berman:2013cli}, the local structure group of the heterotic DFT is given by the maximal compact subgroup of $\ODDG$\cite{Bedoya:2014pma,Coimbra:2014qaa}
\be
\mathbf{O}(1,D-1) \times \ODG \subset \ODDG\,.
\ee
Then we introduce a pair of local orthonormal frame $\{V_{\hM m}\,, \brV_{\hM \hbrm}\}$ corresponding to the $\mathbf{O}(1,D-1) \times \mathbf{O}(D-1,1+\text{dim}\,G)$ respectively, where $\hM$ is an $\ODDG$ vector index, $m$ is an $\OoD$ vector index and $\hbrm$ is an $\ODG$ vector index. They satisfy the following defining properties \cite{Jeon:2011cn},
\be
\ba{llll}
V_{\hM p} V^{\hM}{}_{q}=\eta_{pq}\,,~~~&~~~~
\brV_{\hM\hbrp} \brV^{\hM}{}_{\brq}=\hat\breta_{\hbrp\hat\brq}\,,
\\
V_{\hM p} \brV^{\hM}{}_{\hat\brq}=0\,,~~~&~~~~V_{\hM p}V_{\hN}{}^{p}+\brV_{\hM \hbrp} \brV_{\hN}{}^{\hbrp} = \hat\cJ_{\hM\hN}\,.
\ea
\label{defV}
\ee
where $\eta_{mn}$ and $\hat\breta_{\hbrp\hbrq}$ are $\mathbf{O}(1,D-1)$ and $\mathbf{O}(D-1,1+ \text{dim} \,G)$ metric respectively and $\cJ_{\hM\hN}$ is $\ODDG$ metric. Hence the double-vielbein form  a pair of  rank-two  projections~\cite{Jeon:2010rw},
\be
\ba{ll}
P_{\hM\hN}:= V_{\hM}{}^{p} V_{\hN p}\,,~~~~&~~~~\brP_{\hM\hN}:=\brV_{\hM}{}^{\hbrp}\brV_{\hN\hbrp}\,,
\ea
\ee
and further meet
\be
\ba{llll}
P_{\hM}{}^{\hN} V_{\hN p}= V_{\hM p}\,,~~&~~\brP_{\hM}{}^{\hN}\brV_{\hN\hbrp}=\brV_{\hM\brp}\,,~~&~~\brP_{\hM}{}^{\hN}V_{\hN p}=0\,,~~&~~P_{\hM}{}^{\hN}\brV_{\hN\hbrp}=0\,.
\ea
\ee
As ungauged DFT, the generalized metric is defined by 
\be
\cH_{\hM\hN} = V_{\hM p}V_{\hN}{}^{p} - \brV_{\hM \hbrp} \brV_{\hN}{}^{\hbrp}
\ee

A necessary step to identify the gauged DFT with heterotic supergravity is to fix a parametrization of the double-vielbein in terms of the heterotic supergravity fields. By doing so, it is necessary to decompose $\ODDG$ vector indices $\hM = \{M\,, A\}$ and $\ODG$ vector indices $\hbrm = \{\brm\,, \bra\}$. We shall start by decomposing the $\ODDG$ metric and  $\ODG$ metric as
\be
\cJ_{\hM\hN} = \bpm \cJ_{MN} & 0 \\ 0& \frac{1}{\alphap} {\cal K}_{AB} \epm\,, 
\qquad
\breta_{\hbrm\hbrn} = \bpm \breta_{\brm\brn} & 0 \\ 0 & \kappa_{\bra\brb}\epm\,,
\ee
where $\cJ_{MN}$ is the $\ODD$ metric, $\breta_{\brm\brn}$ is the $\mathbf{O}(D-1,1)$ metric. Also $\cK_{AB}$ and $\kappa_{\bra\brb}$ are defined 
\be
\cK_{A B} = \bpm  \kappa_{\alpha \beta} & 0 \\ 0 & -  \tilde{\kappa}_{\tilde{\alpha} \tilde{\beta}}\epm\,,
\qquad
\kappa_{\bra \brb} = \bpm  \kappa_{ij} & 0 \\ 0 & -  \tilde{\kappa}_{[\brm\brn][\brp\brq]}\epm\,.
\ee
Here we employ the collective adjoint gauge indices $\bra= \{i,\,[\brm\brn]\}$, where $i,j, \cdots$ correspond to heterotic Yang-Mills group and $[\brm\brn], [\brp\brq] \cdots$ correspond to $\mathbf{O}(D-1,1)$ local Lorentz group. The $A,B,\cdots$ indices are pull back of $\bra,\brb,\cdots$ indices by using the collective generator $t_{A}{}^{\bra}$, where $A, B,\cdots = 1~ \text{to}~ (\text{dim}\,G)^2$ and $\bra,\brb, \cdots = 1 ~\rm{to}~ \text{dim}\,G$. (See appendix \ref{appendixA} for further comments about the conventions.) 
Next we decompose the double-vielbein as well
\be
V_{\hat{M}}{}^{m} = \bpm V_{M}{}^{m} \\ V_{A}{}^{m} \epm\,, \qquad 
\brV_{\hat{M}}{}^{\hat{\brm}} = \bpm \brV_{M}{}^{\brm} & \brV_{M}{}^{\bra} \\ \brV_{A}{}^{\brm} & \brV_{A}{}^{\bra} \epm\,.
\label{vielbeindecomp}\ee
Under these decompositions the defining condition of the double-vielbein (\ref{defV}) is then reduced to
\be\ba{ll}
(1)\quad V_{M m} \cJ^{MN} V_{N n} + V_{A m} K^{AB} V_{B n} = \eta_{mn}\,, 
\\
(2)\quad \brV_{M \brm} \cJ^{MN} \brV_{N \brn} + \brV_{A \brm} K^{AB} \brV_{B \brn} = \breta_{\brm\brn}\,,
\\
(3)\quad \brV_{M \brm} \cJ^{MN} \brV_{N \bra} + \brV_{A \brm} K^{AB} \brV_{B \bra} = 0\,, 
\\ 
(4)\quad \brV_{M \bra} \cJ^{MN} \brV_{N \brb} + \brV_{A \bra} K^{AB} \brV_{B \brb} = \kappa_{\bra\brb}\,, 
\\
(5)\quad V_{M m} \cJ^{MN} \brV_{N \brn} + V_{A m} K^{AB} \brV_{B \brn} = 0\,,
\\
(6)\quad V_{M m} \cJ^{MN} \brV_{N \bra} + V_{A m} K^{AB} \brV_{B \bra} = 0\,,
\\
(7)\quad V_{M}{}^{m} V_{N m} + \brV_{M}{}^{\brm} \brV_{N \brm} + \brV_{M}{}^{\bra} \brV_{N \bra} = \cJ_{MN}\,,
\\
(8)\quad V_{M}{}^{m} V_{A m} + \brV_{M}{}^{\brm} \brV_{A \brm} + \brV_{M}{}^{\bra} \brV_{A \bra} = 0\,,
\\
(9)\quad V_{A}{}^{m} V_{B m} + \brV_{A}{}^{\brm} \brV_{B \brm} + \brV_{A}{}^{\bra} \brV_{B \bra} = K_{AB}
\ea\label{DecompDefiningProp}\ee
We then construct a parametrization satisfying the defining properties (\ref{DecompDefiningProp}) and assuming the upper half block of $V_{M}{}^{m}$ and $\brV_{M}{}^{\brm}$ are non-degenerated 
\begin{equation}
V_{M}{}^m = \tfrac{1}{\sqrt{2}} \begin{pmatrix} (e^{-1})^{\mu m} \\ e_\mu{}^{m} + B'_{\mu\nu} (e^{-1})^{\nu m}  \end{pmatrix}\,, \qquad
V_{A}{}^{m} = \cA_{M A} V^{Mm} \,, 
\label{parametrizationV}
\end{equation}
and for $\brV_{\hM}{}^{\hbrm}$ 
\begin{equation}
\ba{ll}
\brV_{M}{}^{\brm} = \tfrac{1}{\sqrt{2}} \begin{pmatrix} (\bre^{-1})^{\mu \brm} \\ \bre_\mu{}^{\brm} + B'_{\mu\nu} (\bre^{-1})^{\nu \brm}  \end{pmatrix}\,, \qquad
\brV_{A}{}^{\brm} = \cA_{M A} \brV^{M\brm} \,, 
\\
\brV_{M}{}^{\bra} =-\sqrt{\alphap} \cA_{M}{}^{\bra}\,, \qquad \brV_{A}{}^{\bra} = \tfrac{1}{\sqrt{\alphap}} (t^{\bra})_{A} \,,
\ea\label{parametrizationbrV}
\end{equation}
where $e_{\mu}{}^{m}$ and $\bre_{\mu}{}^{\brm}$ are two copies of the $D$-dimensional vielbein corresponding to the same metric $g_{\mu\nu}$
\be
e_{\mu}^{m} e_{\nu}{}^{n} \eta_{mn} = - \bre_{\mu}{}^{\brm} \bre_{\nu}{}^{\brn} \breta_{\brm\brn} = g_{\mu\nu}\,.
\ee
and  $\cA_{M}{}^{\bra}$ and  $B'_{\mu\nu}$ are defined as
\be\ba{ll}
\cA_{M}{}^{\bra} := \cA_{M}{}^{A} t_{A}{}^{\bra}\,,
\\
B'_{\mu\nu} := B_{\mu\nu} - \tfrac{1}{2} \alphap \cA_{\mu}{}^{\bra} \cA_{\nu \bra}\,,
\ea\ee 
Note that the double-vielbein under the parametrization is identical with the frame field in generalized geometry in a local coordinate patch \cite{Coimbra:2014qaa}.
Here $\cA_{M}{}^{\bra}$ should be identified as a gauge field for $G$ by its transformation property. Since $\cA_{M}{}^{\bra}$ is parametrized as
\be
\cA_{M}{}^{\bra} = \bpm0\\\cA_{\mu}{}^{\bra}\epm\,.
\label{derivIndexValued}\ee
it is so called the {\it derivative index valued field} \cite{Park:2013mpa,Lee:2013hma}.

The gauge field $\cA_{M}{}^{\bra}$ consists of two gauge fields for $\mathbf{SO}(32)$ or $E_8 \times E_8$ heterotic Yang-Mills symmetry and $\mathbf{O}(D-1,1)$ local Lorentz transformation
\be
\cA_{M}{}^{\bra} = \{\cA_{M}{}^{i}\,, \cA_{M}{}^{[\brm\brn]}\}
\ee
Since these two gauge fields appear symmetrically in the action and supersymmetry transformation \cite{Bergshoeff:1989de,Bedoya:2014pma,Coimbra:2014qaa}, we will use a combined form $\cA_{M}{}^{\bra}$ unless we have to distinguish them. 
 
Note that the $\brV^{\mu\bra} = 0$ and $\brV_{A}{}^{\bra} = \tfrac{1}{\sqrt{\alphap}} t_{A}{}^{\bra}$ are not preserved under the $\ODG$ local Lorentz group. To preserve the parametrization the $\ODG$ transformation should be broken to $\mathbf{O}(D-1,1)$ subgroup. Therefore, after parametrization, the local Lorentz group is reduced to $\OoD\times\mathbf{O}(D-1,1)$ as ordinary DFT.

As shown in \cite{Bedoya:2014pma,Coimbra:2014qaa}, the gauge field $\cA_{M [\brm\brn]}$ for $\mathbf{O}(D-1,1)$  local Lorentz group should be identified with DFT spin-connection $\brPhi_{M\brm\brn}$ (\ref{spinconnections}) in order to describe $\alpha'$-correction. However the defining properties (\ref{DecompDefiningProp}) are not enough to fix the explicit form of $\cA_{M[\brm\brn]}$, so the identification should be imposed by hand.
Nevertheless, there is still ambiguity in determining the explicit form of $\cA_{M[\brm\brn]}$. The identification, $\cA_{M [\brm\brn]} = \brPhi_{M \brm\brn}$, is inconsistent with SUSY transformation, and it will be discussed later after introducing fermionic sector. 
It is also important to note that $\cA_{M [\brm\brn]}$ is a composite field, thus it does not give additional degrees of freedom.

Under the previous parametrization, the projection operators are parametrized 
\be
P_{\hM\hN} =  V_{\hM}{}^{m} V_{\hN m}\,, \qquad 
\brP_{\hM\hN} = \brV_{\hM}{}^{\brm} \brV_{\hn \brm} + \brV_{\hM}{}^{\bra} \brV_{\hn \bra}\,,
\ee
where 
\be
P_{\hM\hN} = 
	\half \bpm
		g^{\mu\nu} & \delta^{\mu}{}_{\nu} + g^{\mu\rho} (B')^{t}_{\rho\nu} &  g^{\mu\nu} \cA_{\nu B}\\
		\delta_{\mu}{}^{\nu} + B'_{\mu\rho} g^{\rho \nu}& \quad g_{\mu\nu} - \alphap \cA_{\mu}{}^{\bra} \cA_{\nu\bra} 
		+ B'_{\mu\rho} g^{\rho\sigma} (B')^t_{\sigma\nu} 
		&\quad \cA_{\mu B} + B'_{\mu\nu} g^{\nu\rho} \cA_{\rho B}\\
		(\cA)^t_{A \mu} g^{\mu\nu} & (\cA)^t_{A \mu} + (\cA)^t_{A\rho} g^{\rho\sigma} (B')^t_{\sigma\nu} & 
		g^{\mu\nu} \cA_{\mu A} \cA_{\nu B}
	\epm\,,
\ee
and
\be
\brP_{\hM\hN} = 
	\half \bpm
		- g^{\mu\nu} & \delta^{\mu}{}_{\nu} - g^{\mu\rho} (B')^{t}_{\rho\nu} & -g^{\mu\nu} \cA_{\nu B}
		\\
		\delta_{\mu}{}^{\nu} - B'_{\mu\rho} g^{\rho \nu}& \quad - g_{\mu\nu}  + \alphap \cA_{\mu}{}^{\bra} \cA_{\nu\bra} 
		- B'_{\mu\rho} g^{\rho\sigma} (B')^t_{\sigma\nu}
		&\quad -\cA_{\mu B} - B'_{\mu\nu} g^{\nu\rho} \cA_{\rho B}
		\\
		- (\cA)^t_{A \mu} g^{\mu\nu} & - (\cA)^t_{A \mu} - (\cA)^t_{A\rho} g^{\rho\sigma} (B')^t_{\sigma\nu} & 
		- g^{\mu\nu} \cA_{\mu A} \cA_{\nu B} + \frac{2}{\alphap}t_{A}{}^{\bra} t_{B \bra}
	\epm\,,
\ee
Then, one can show that $\cJ_{\hM\hN} = P_{\hM\hN} + \brP_{\hM\hN}$ and the well-known generalized metric (see also \cite{Blumenhagen:2014iua} for non-geometric parametrization) is reproduced from $\cH_{\hM\hN} = P_{\hM\hN} - \brP_{\hM\hN}$
\be
\cH_{\hM\hN} = 
	\bpm
		g^{\mu\nu} & g^{\mu\rho} (B')^t_{\rho\nu} & g^{\mu\nu} \cA_{\mu B}
		\\
		B'_{\mu\rho} g^{\rho\nu} & \quad g_{\mu\nu} - \alphap \cA_{\mu}{}^{\bra} \cA_{\nu \bra} 
		+ B'_{\mu\rho} g^{\rho\sigma} (B')_{\sigma\nu} &\quad \cA_{\mu B} + B'_{\mu\nu} g^{\nu\rho} \cA_{\rho B}
		\\
		(\cA)^t_{A \mu} g^{\mu\nu} & (\cA)^t_{A \mu} + (\cA)^t_{A\rho} g^{\rho\sigma} (B')^t_{\sigma\nu} & 
		g^{\mu\nu} \cA_{\mu A} \cA_{\nu B} - \frac{1}{\alphap}t_{A}{}^{\bra} t_{B \bra}
	\epm\,.
\ee

After the parametrization, all the symmetries of the $\ODDG$ gauged DFT are partially broken or modified. In the rest of this section we will consider the bosonic symmetries and their compensating local Lorentz transformation which sustains the parametrization.

\subsection{Gauge transformations}
In the gauged DFT, the generalized Lie derivative in ungauged DFT is replaced by the twisted generalized Lie derivative which includes Yang-Mills gauge symmetry. For the $\ODDG$ double vielbeins, the twisted generalized Lie derivative is defined \cite{Hohm:2011ex,Grana:2012rr}
\be
\hat{\cL}_{\hX} V_{\hM}{}^{m} = 
	\hcLo_{\hX} V_{\hM}{}^{m} 
	- \tfrac{1}{\sqrt{\alphap}} f_{\hM\hP\hQ} {X}^{\hP} V^{\hQ m}\,,
\label{generalizedLieDeriv}\ee
where $\hcLo$ is the ordinary generalized Lie derivative,
\be
\hcLo_{\hX} V_{\hM}{}^{m}=  X^{\hP} \partial_{\hP} V_{\hM}{}^{m} + \big(\partial_{\hM} X^{\hP} - \partial^{\hP} X_{\hM}\big) V_{\hP}{}^{m}\,,
\ee
and $f_{\hM\hP\hQ}$ is the structure constant for the gauge group $G$ in $\ODDG$ covariant manner. The section condition also known as the strong constraint is given by: 
\be
\partial_{\hM} \partial^{\hM} \Phi = 0\,, ~~~~~~~~~~~\partial_{\hM} \Phi_{1} \, \partial^{\hM} \Phi_{2} = 0\,
\ee
The structure constants $f_{\hM\hN\hP}$ should then satisfy the Jacobi identity,
\be
f_{\hM[\hN}{}^{\hP} f_{|\hP|\hQ\hR]} = 0\,.
\ee 
It is also convenient to impose an orthogonality condition on the structure constants $f_{\hM\hN\hP}$ 
\be
f_{\hM\hN\hP} \, \partial^{\hM} X = 0\,,
\ee
This means the gauge symmetry will be orthogonal to the ordinary generalized Lie derivative.
The gauge parameter ${X}{}^{\hM}$ consists of diffeomorphism parameter $\xi^{\mu}$, one-form gauge parameter $\Lambda_{\mu}$ for Kalb-Ramond field $B_{\mu\nu}$ and Yang-Mills gauge parameter $\lambda^{A}$ for gauge group $G$ as
\be
{X}^{\hM} = \{\xi^{\mu}\,, \Lambda_{\mu} \,, \sqrt{\alphap} \lambda^{A} \}\,.
\ee

However, the twisted generalized Lie derivative does not sustain the previous double-vielbein parametrization. For example, if we transform the constant component $\brV_{A}{}^{\bra}$, then it doesn't remain as a constant,
\be
\hcL_{\hX} \brV_{A}{}^{\bra} = 
	f_{A}{}^{B}{}_{C}  \brV_{B}{}^{\bra} \lambda^{C}\,.
\label{localLorentzParameter}\ee
To overcome this problem, we modify the twisted generalized Lie derivative by adding a compensating Lorentz transformation which cancels the unwanted terms as in \cite{Bedoya:2014pma}
\be\ba{ll}
\delta_{X} \brV_{M}{}^{\bra} &=  
	X^{N} \partial_{N} \brV_{M}{}^{\bra}
	+ \left(\partial_{M}X^{N} - \partial^{N} X_{M} \right) \brV_{N}{}^{\bra} 
	+ \partial_{M} X^{A} \brV_{A}{}^{\bra}
	+ \Lambda^{\bra}{}_{\brb} \brV_{M}{}^{\brb}\,,
\\
\delta_{X} \brV_{A}{}^{\bra} & = 
	- f_{ABC} X^{B} \brV^{C \bra} 
	+ \Lambda^{\bra}{}_{\brb} \brV_{A}{}^{\brb}\,,
\ea\ee
where $\Lambda^{\bra}{}_{\brb}\in \mathbf{O}(\text{dim} \, G)$ is taken to be
\be
\Lambda^{\bra}{}_{\brb} = \sqrt{\alpha'} f^{\bra}{}_{\brb\brc} \lambda^{\brc}\,.
\ee

Then we reproduce the gauge transformations for component fields of the double-vielbein, which have been constructed in \cite{Bedoya:2014pma} 
\be\ba{ll}
\delta e_{\mu}{}^{m} &= \xi^{\nu} \partial_{\nu} e_{\mu}{}^{m} + \partial_{\mu} \xi^{\nu} e_{\nu}{}^{m}\,,
\\
\delta \cA_{\mu}{}^{\bra} & = \xi^{\nu} \partial_{\nu} \cA_{\mu}{}^{\bra} + \partial_{\mu} \xi^{\nu} \cA_{\nu}{}^{\bra} - \partial_{\mu} \lambda^{\bra} + f^{\bra}{}_{\brb\brc} \cA_{\mu}{}^{\brb} \lambda^{\brc}\,,
\\
\delta B_{\mu\nu} & =  \xi^{\rho} \partial_{\rho} B_{\mu\nu} + \partial_{\mu} \xi^{\rho} B_{\rho \nu} + \partial_{\nu} \xi^{\rho} B_{\mu \rho} + 2 \partial_{[\mu} \Lambda_{\nu]} + \alphap \partial_{[\mu} \lambda^{\bra} \cA_{\nu] \bra} \,.
\ea
\ee

\subsection{$\ODD$ transformation}
We now turn to the $\ODDG$ global duality symmetry. Before the parametrization, the double-vielbein is an $\ODDG$ vector which transform as
\be
\delta_h V_{\hM}{}^{m} = h_{\hM}{}^{\hN} V_{\hN}{}^{m} \,, \qquad  \delta_h \brV_{\hM}{}^{\hbrm} = h_{\hM}{}^{\hN} \brV_{\hN}{}^{\hbrm}\,,
\label{ODDGtransf}\ee
where $h_{\hM}{}^{\hN} \in \ODDG$. As shown in  \cite{Hohm:2011ex}, $\ODDG$ symmetry is broken to the $\ODD$ subgroup due to the parametrization of $f_{\hM\hN\hP}$. However the $\ODD$ symmetry should be modified by the parametrization of double-vielbein by introducing a compensating local Lorentz transformation. In here, we will consider infinitesimal $\ODD$ transformation. 

To construct the infinitesimal $\ODD$ transformation, we decompose the double-vielbein transformation as follows:
\be
\bpm\delta_{h} V_{M}{}^{m} \\ \delta_{h} V_{A}{}^{m} \epm 
	= \bpm h_{M}{}^{N} & h_{M}{}^{B} \\  h_{A}{}^{N} & h_{A}{}^{B} \epm \bpm V_{N}{}^{m} \\  V_{B}{}^{m} \epm\,,
\label{ODDtransfV}\ee
and
\be
\bpm\delta_{h} \brV_{M}{}^{\brm} & \delta_{h} \brV_{M}{}^{\bra} \\ 
	\delta_{h} \brV_{A}{}^{\brm} & \delta_{h} \brV_{A}{}^{\bra} & \epm 
		= \bpm h_{M}{}^{N} & h_{M}{}^{B} \\  h_{A}{}^{N} & h_{A}{}^{B} \epm 
		\bpm \brV_{N}{}^{\brm} & \brV_{N}{}^{\bra} \\ \brV_{B}{}^{\brm} & \brV_{B}{}^{\bra}\epm\,,
\label{ODDtransfbrV}\ee
where $h_{MN}$ and $h_{A B}$ are antisymmetric matrices. The $\mathbf{SO}(D,D)$ algebra element $h_{M}{}^{N}$ admits further decomposition as
\be
h_{MN} = 
	\bpm \alpha^{\mu\sigma} & -(\beta^t)^{\mu}{}_{\rho} 
	\\ 
	\beta_{\nu}{}^{\sigma} & \gamma_{\nu\rho}\epm\,,
		\qquad \text{or} \qquad
h_{M}{}^{N} = 
	\bpm -(\beta^t)^{\mu}{}_{\rho}& \alpha^{\mu\sigma} 
	\\ 
	\gamma_{\nu\rho} & \beta_{\nu}{}^{\sigma}\epm\,,
\ee 
where $\alpha^{\mu\nu}$ and $\gamma_{\mu\nu}$ are antisymmetric parameters
\be
\alpha^{\mu\nu} = - \alpha^{\nu\mu} \,, \qquad \gamma_{\mu\nu} = - \gamma_{\nu\mu}\,.
\ee

Having the parametrization of double-vielbein, there is a consistency condition that must be obeyed by the $\ODD$ transformation. Since $V_{\hM}{}^{m}$ and $\brV_{\hM}{}^{\hbrm}$ are parametrized in terms of the same heterotic supergravity fields, in the component field language, (\ref{ODDtransfV}) and (\ref{ODDtransfbrV}) should be consistent to each other. For instance, we can obtain $\delta_h B_{\mu\nu}$ or $\delta_h \cA_{\mu A}$ from each of (\ref{ODDtransfV}) and (\ref{ODDtransfbrV}) independently, then consistency requires these results should be identical to each other.

First, the $\ODD$ transformation of $V_{\hM}{}^{m}$ (\ref{ODDtransfV}) leads 
\be\ba{rl}
\delta_{h} e_{\mu}{}^{m}  &= 
	\beta_{\mu}{}^{\nu} e_{\nu}{}^{m} 
	+ ( g_{\mu\nu} - B_{\mu\nu}) \alpha^{\nu \rho} e_{\rho}{}^{m} 
	- \half \alphap \cA_{\mu}{}^{\bra} \cA_{\nu\bra}\alpha^{\nu\rho} e_{\rho}^{m}\,,
\\
\delta_{h} B_{\mu\nu} &= 
	\gamma_{\mu\nu}  
	+ 2B_{[\mu|\rho|} (\beta^t)^{\rho}{}_{\nu]} 
	- g_{\mu \rho} \alpha^{\rho\sigma} g_{\sigma\nu}  
	- B_{\mu \rho} \alpha^{\rho\sigma} B_{\sigma\nu} \,,
\\&\quad 
	+ \alphap \cA_{[\mu}{}^{\bra} \cA_{|\rho \bra} \alpha^{\rho \sigma} g_{\sigma | \nu]} 
	- \tfrac{1}{4} \alphap^2 \cA_{\mu}{}^{\bra} \cA_{\nu}{}^{\brb} \cA_{\rho \bra}\alpha^{\rho\sigma} \cA_{\sigma \brb}	\,,
\\
\delta_{h} \cA_{\mu A}& = 
	\beta_{\mu}{}^{\rho} \cA_{\rho A} 
	+ g_{\mu\nu} \alpha^{\nu\rho} \cA_{\rho A} 
	- B_{\mu\nu} \alpha^{\nu\rho} \cA_{\rho A} 
	- \half\alpha' \cA_{\mu}{}^{\bra} \cA_{\rho\bra} \alpha^{\rho\sigma} \cA_{\sigma A} 
\ea\label{componentODD}\ee
On the other hand, if we evaluate $\delta_{h}\brV_{\hM}{}^{\brm}$ part only by using the parametrization of $\brV_{\hM}{}^{\brm}$  (\ref{parametrizationbrV}), then we get inconsistent result with (\ref{componentODD}). Moreover, even though $\brV_{M}{}^{\bra}$ is parametrized as a derivative index valued vector, namely $\brV^{\mu \bra} = 0$, but under the naive $\ODDG$ transformation (\ref{ODDtransfbrV})
\be
\delta_{h} \brV^{\mu\bra} = - \sqrt{\alphap} \alpha^{\mu\nu} \cA_{\nu}{}^{\bra} \neq 0\,.
\ee 

These problems can be solved by adding a compensating $\ODG$ local Lorentz transformation on the $\ODG$ vector indices as follows
\be
\tilde{\delta}_h \brV_{\hM}{}^{\hbrm} := \bpm
	h_{M}{}^{N}{} \brV_{N}{}^{\hbrm} + \brV_{N}{}^{\bra} \Lambda_{\bra}{}^{\hbrm}~ & 
	\\
	h_{A}{}^{M} \brV_{M}{}^{\hbrm} + \brV_{A}{}^{\bra} \Lambda_{\bra}{}^{\hbrm}
	\epm\,,
\label{modODDGbrV}\ee
where 
\be\ba{lll}
\Lambda_{\brm}{}^{\bra} =& 
	\sqrt{2 \alphap} \bre_{\mu\brm} \alpha^{\mu\nu} \cA_{\nu}{}^{\bra}\,,
\qquad
\Lambda_{\brb}{}^{\bra} =& -\alphap \cA_{\mu \brb} \alpha^{\mu\nu} \cA_{\nu}{}^{\bra}\,.
\ea\ee
If we evaluate $\tilde{\delta}_{h} \brV_{M}{}^{\brm}$ and $\tilde{\delta}_{h} \brV_{A}{}^{\brm}$¸ then one can show that these are consistent with (\ref{ODDtransfV}) in the component field level. Moreover we can show that $\tilde{\delta}_h V_{A}{}^{\bra}$ is vanished, and it is consistent with the fact that $\brV_{A}{}^{\bra}$ is the structure constant of $G$. 
\subsection{Connection}
We now introduce geometrical quantities defined in gauged DFT to describe the dynamics and supersymmetry of heterotic DFT. 

As for the covariant differential operator of the heterotic DFT, we present a covariant derivative which can be applied to any arbitrary $\ODDG$, $\Spin(1,D-1)$ and $\Spin(D-1,1+\rm{dim}~G)$ representations as follows 
\be
\hat\cD_{\hM} := \partial_{\hM}  + \Gamma_{\hM} + \Phi_{\hM} + \brPhi_{\hM}\, . 
\label{MasterDerivative}\ee
where $\Phi_{Mmn}$ and $\brPhi_{M\brm\brn}$ are spin-connections and $\Gamma_{\hM\hN\hP}$ is semi-covariant connection which are constructed in gauged DFT \cite{Berman:2013cli}
\be  
\Gamma_{\hP\hM\hN} = 
	\Gammao_{\hP\hM\hN} 
	+ \left(\delta_{P}{}^{\hQ} P_{\hM}{}^{\hR} P_{\hN}{}^{\hS} 
	+ \delta_{\hP}{}^{\hQ} \brP_{\hM}{}^{\hR} \brP_{\hN}{}^{\hS} \right) f_{\hQ\hR\hS} 
	-\tfrac{2}{3} \left(\cP + \bar{\cP}\right)_{\hP\hM\hN}{}^{\hQ\hR\hS} f_{\hQ\hR\hS}\,.
\label{conn}\ee
where $\Gammao_{PMN}$ is the connection for ordinary DFT \cite{Jeon:2011cn},
\be
\begin{array}{ll}
\Gammao_{\hP\hM\hN}  = & 
	2(P\partial_{\hP} P \brP )_{[\hM\hN]} 
	+ 2 (\brP_{[\hM}{}^{\hQ} \brP_{\hN]}{}^{\hR} 
	- P_{[\hM}{}^{\hQ} P_{\hN]}{}^{\hR} ) \partial_{\hQ} P_{\hR \hP} 
\\& 
	- \tfrac{4}{D-1} \big(\brP_{P[\hM} \brP_{\hN]}{}^{\hQ} + P_{\hP[\hM} P_{\hN]}{}^{\hQ}) 
		\big(\partial_{\hQ}d + (P\partial^{\hR} P \brP\big)_{[\hR\hQ]}\big)\,,
\end{array}
\label{oldconn}
\ee 
and $\cP_{\hP\hM\hN}{}^{\hQ\hR\hS}$ and $\bar{\cP}_{\hP\hM\hN}{}^{\hQ\hR\hS}$ are rank-six projection operators 
\be
\ba{ll}
\cP_{\hP\hM\hN}{}^{\hS\hQ\hR}:=& P_{\hP}{}^{\hS}P_{[\hM}{}^{[\hQ}P_{\hN]}{}^{\hR]}+\tfrac{2}{D-1}P_{\hP[\hM}P_{\hN]}{}^{[\hQ}P^{\hR]\hS}\,,
\\
\bcP_{\hP\hM\hN}{}^{\hS\hQ\hR}:=&\brP_{\hP}{}^{\hS}\brP_{[\hM}{}^{[\hQ}\brP_{\hN]}{}^{\hR]}+\frac{2}{D-1}\brP_{\hP[\hM}\brP_{\hN]}{}^{[\hQ}\brP^{\hR]\hS}\,,
\ea\ee
which are symmetric and traceless,
\be
\ba{ll}
{\cP_{\hP\hM\hN\hQ\hR\hS}=\cP_{\hQ\hR\hS\hP\hM\hN}=\cP_{\hP[\hM\hN]\hQ[\hR\hS]}\,,}
~~&~~
	{\bcP_{\hP\hM\hN\hQ\hR\hS}=\bcP_{\hQ\hR\hS\hP\hM\hN}=\bcP_{\hP[\hM\hN]\hQ[\hR\hS]}\,,} 
\\
{\cP^{\hP}{}_{\hP\hM\hQ\hR\hS}=0\,,~~~~\,P^{\hP\hM}\cP_{\hP\hM\hN\hQ\hR\hS}=0\,,}
~~&~~
	{\bcP^{\hP}{}_{\hP\hM\hQ\hR\hS}=0\,,~~~~\,\brP^{\hP\hM}\bcP_{\hP\hM\hN\hQ\hR\hS}=0\,.}
\ea
\label{symP6}
\ee
Here the superscript ‘0’ indicates a quantity defined in the ungauged DFT.

To determine the spin-connections in (\ref{MasterDerivative}), we impose the double-vielbein compatibility condition
\be
\hat\cD_{\hM} V_{\hN m} = 0\,, ~~~~~~~~ \hat\cD_{\hM} \brV_{\hN\brm} = 0\,,
\label{compv}
\ee
and for the metric of $\Spin(1,D-1)$ and $\Spin(D-1,1+\rm{dim}\,G)$, $\eta_{mn}$ and $\breta_{\hbrm\hbrn}$ respectively,
\be
\hat\cD_{\hM} \eta_{mn} =  0\,, ~~~~~~~~~~ \hat\cD_{\hM} \breta_{\hbrm\hbrn} = 0\,.
\ee
From the compatibility of $\eta_{mn}$ and $\breta_{\brm\brn}$, we can deduce that the spin-connections are antisymmetric,
\be
\Phi_{\hM mn} = \Phi_{\hM [mn]}\,, ~~~~~~~~ \brPhi_{\hM\hbrm\hbrn} = \brPhi_{M[\hbrm\hbrn]}\,.
\ee
In addition, because of the double-vielbein compatibility condition (\ref{compv}), the spin-connections may be determined in terms of the double-vielbeins as follows,
\be
\Phi_{\hM mn} = V^{\hN}{}_{m} \hat\nabla_{\hM} V_{\hN n}\,, ~~~~~~~~ \brPhi_{\hM \hbrm\hbrn} = \brV^{\hN}{}_{\hbrm} \nabla_{\hM} \brV_{\hN \hbrn}\,,
\label{spinconn}\ee 
where $\hat\nabla_{\hM}$ is the covariant derivative which acts on the $\ODDG$ vector indices
\be
\hat\nabla_{\hM} T_{\hN} = \partial_{\hM} T_{\hN} + \Gamma_{\hM\hN}{}^{\hP} T_{\hP}\,.
\ee
Crucially, we can then form fully covariant quantities with projection operators or double-vielbeins as shown below:
\be
\ba{llllll}
\brV^{\hM}{}_{\hbrp}\Phi_{\hM mn}\,,~&~V^{\hM}{}_{p}\brPhi_{\hM \hbrm\hbrn}\,,~&~\Phi_{\hM[pq}V^{\hM}{}_{r]}\,,~&~
\brPhi_{\hM[\hbrp\hbrq}\hbrV^{\hM}{}_{\hat{\bar{r}}]}\,,~&~\Phi_{\hM pq} V^{\hM p}\,,~&~\brPhi_{\hM\hbrp\hbrq} \brV^{\hM\hbrp}\,.
\ea
\label{covPhi}
\ee 
After the parametrization, the previous covariant spin-connections are decomposed naturally
\be\ba{ll}
&\Phi_{\brp mn}\,,\qquad \Phi_{\bra mn}\,,\qquad \Phi_{[pmn]}\,,\qquad \Phi^{p}{}_{pm}\,,
\\
&\brPhi_{p \brm\brn}\,,\qquad \brPhi_{p \brm\bra}\,,\qquad \brPhi_{p \bra\brb}\,,\qquad \brPhi_{[\brp\brm\brn]}\,,\qquad \brPhi_{[\brp\brm\bra]}\,,\qquad \brPhi_{[\brp\bra\brb]}\,,
\\ 
&\brPhi_{[\bra\brb\brc]}\,,\qquad \brPhi^{\hbrp}{}_{\hbrp \brm}\,,\qquad \brPhi^{\hbrp}{}_{\hbrp \bra}\,.
\ea\ee
In (\ref{spinconnections}), we present explicit form of the spin-connections in terms of heterotic supergravity fields. These will be a building block that the formalism uses. Various covariant quantities can be generated by using these spin-connections and their derivatives \cite{Berman:2013cli}.

\subsection{Curvature}
Let us turn to semi-covariant curvature tensor $S_{\hM\hN\hP\hQ}$ which is defined as
\be
S_{\hM\hN\hP\hQ} = \half \big(R_{\hM\hN\hP\hQ} + R_{\hP\hQ\hM\hN} - \Gamma^{\hR}{}_{\hM\hN} \Gamma_{\hR\hP\hQ} \big)\,,
\label{curvature}\ee
where $R_{\hM\hN\hP\hQ}$ is defined from the standard commutator of the covariant derivatives
\be
R_{\hM\hN\hP\hQ} = \partial_{\hM}\Gamma_{\hN\hP\hQ} - \partial_{\hN}\Gamma_{\hM\hP\hQ} + \Gamma_{\hM\hP}{}^{\hR} \Gamma_{\hN\hR\hQ} -  \Gamma_{\hN\hP}{}^{\hR} \Gamma_{\hM\hR\hQ} + f_{\hR\hM\hN} \Gamma^{\hR}{}_{\hP\hQ}\,.
\ee
The generalized curvature scalar is defined by contraction of $S_{\hM\hN\hP\hQ}$ with the projection operators
\be\ba{rl}
S := &P^{\hM\hN} P^{\hP\hQ} S_{\hM\hP\hN\hQ} 
\\
	=& 
	2 \partial^{m} \Phi^{n}{}_{mn} 
	- \Phi^{m}{}_{m}{}^{p} \Phi^{n}{}_{np} 
	-\frac{3}{2} \Phi^{[mnp]} \Phi_{mnp} 
	-\half \Phi^{\brp mn} \Phi_{\brp mn} 
	-\half \Phi^{\bra mn} \Phi_{\bra mn} 
\\&
	 - f_{p mn } \Phi^{pmn} 
	 - f_{\brp mn } \Phi^{\brp mn}
	 - f_{\bra mn } \Phi^{\bra mn}\,.
\ea\ee
Thus it provides a scalar invariant under the all kinds of bosonic symmetries and gives our bosonic Lagrangian in a compact form 
\be
\cS_{B} = \int e^{-2d} 2S\,.
\ee
From the variation of the heterotic DFT action with respect to double-vielbein, the corresponding generalized Ricci tensor is defined as
\be\ba{rl}
S_{m\hbrn}  :=& 
	V_{m}{}^{\hM} \brV_{\hbrm}{}^{\hN} S_{\hM\hP\hN\hQ} P^{\hP\hQ}\,,
\\
	=& \half \big(\partial_{m} \brPhi^{\hbrp}{}_{\brn \hbrp} 
	- \partial^{\brp}\brPhi_{m\hbrn\hbrp} 
	+ \brPhi_{m\hbrn}{}^{\hbrp} \brPhi^{\hbrq}{}_{\hbrp \hbrq} 
	+ \Phi^{\hbrp}{}_{q m} \Phi^{q}{}_{\hbrn\hbrp} \big)\,.
\ea\label{GenerlizedRicciTensor}\ee

Now let's consider Bianchi identity for curvature
\be
S_{[\hM\hN\hP]\hQ} = 0\,.
\ee
If we pull back to the frame indices by using double-vielbein, we have 
\be\ba{ll}
S_{[mnp]q} = 
	4 \partial_{[m} \Phi_{npq]} + \Phi_{\brr[mn} \Phi^{\brr}{}_{pq]} + \Phi_{\bra[mn} \Phi^{\bra}{}_{pq]}
	+9 \Phi^{(A)}{}_{r[mn} \Phi^{(A)}{}^{r}{}_{pq]}\,,
\\
S_{[mnp]\bra} = \half \big(
	\partial_{[p}\Phi_{|\bra|mn]} 
	- \Phi_{\brp[mn} \brPhi_{p]}{}^{\brp}{}_{\bra} 
	- \Phi_{\brb[mn} \brPhi_{p]}{}^{\brb}{}_{\bra}
	+ 3 \Phi_{\bra}{}^{q}{}_{[p} \Phi^{(A)}{}_{mn]q}
	\big)
\ea\label{BI0}\ee
where $\Phi^{(A)}{}_{mnp} = \tfrac{1}{3} \Phi_{[mnp]}$.
If we substitute the explicit form of spin-connections, then (\ref{BI0}) is reduced to the anomaly cancelation condition and Jacobi identity for the Yang-Mills field strength as we desired
\be\ba{ll}
S_{[mnp]q} = 
	e_{m}{}^{\mu}e_{n}{}^{\nu}e_{p}{}^{\rho}e_{q}{}^{\sigma} \big(
		\frac{1}{3}\partial_{[\mu} H_{\nu\rho\sigma]} 
		- \frac{1}{4} \Tr(F\wedge F - R^{(-)}\wedge R^{(-)})_{[\mu\nu\rho\sigma]} 
		\big)\,,
\\
S_{[mnp]\bra} = e_{m}{}^{\mu}e_{n}{}^{\nu}e_{p}{}^{\rho} \big(
	\frac{1}{2\sqrt2} D_{[\mu} (\cF_{\nu\rho]})_{\bra}
	\big)\,.
\ea\label{BI}\ee
This feature is exactly same as \cite{Bedoya:2014pma, Coimbra:2014qaa}, and it shows the consistency of the semi-covariant formulation for heterotic DFT.

\section{Supersymmetry in leading order $\alphap$-correction}
In this section we consider supersymmetry in heterotic DFT with leading order $\alphap$-corrections based on supersymmetric gauged DFT \cite{Berman:2013cli}. We also consider the relation with the generalized geometry result \cite{Coimbra:2014qaa}. 

As shown in the previous section, the bosonic sector consists of DFT-dilaton, $d$, and double-vielbeins, $V_{\hM m}$, $\brV_{\hM\brm}$. Meanwhile the fermionic sector is determined by the supersymmetry. Since heterotic DFT admits $N=1$ supersymmetry, the fermonic degrees of freedom are given by one kind of gravitino $(\psi_{\brm})^{\alpha}$, gauginos $(\psi_{\bra})^{\alpha}$ and the dilatino $(\rho)^{\alpha}$, where $\alpha\,, \beta\,, \cdots$ represent the spinor representation of $\Spin(1,9)$. Here we employ a collective notation for the gauginos as the gauge field $\cA_{M\bra}$ 
\be
\psi_{\bra} = \{\chi^{i}\,, \psi_{[\brm\brn]} \}\,,
\ee 
where $\psi_{[\brm\brn]}$ is so called {\it gravitino curvature} \cite{Bergshoeff:1989de} which is defined in terms of heterotic DFT variables
\be\ba{rl}
\psi_{[\brm\brn]} := & 2 \partial_{[\brm} \psi_{\brn]} + \half \Phi_{[\brm |pq|} \gamma^{pq} \psi_{\brn]}  + 2 \brPhi_{[\brm\brn]}{}^{\brp} \psi_{\brp} + \brPhi^{\brp}{}_{\brm\brn} \psi_{\brp} - \cA^{\brp}{}_{\brm\brn} \psi_{\brp}\,,
\\
=&  2 \cD_{[\brm} \psi_{\brn]} + \brPhi^{\brp}{}_{\brm\brn} \psi_{\brp} - \cA^{\brp}{}_{\brm\brn} \psi_{\brp}\,.
\ea\label{GravitinoCurvature}\ee
Here $\cD_{\brm}$ is a covariant derivative for $\Spin(1,9)$ and $\Spin(9,1)$ vector representation, for instance arbitrary vectors $T_{m}$ and $T_{\brm}$
\be
\cD_{\brm} T_{m} = 
	\partial_{\brm} T_{m} 
	+ \Phi_{\brm n}{}^{p} T_{p}\,,
\qquad 
\cD_{m} T_{\brn} = 
	\partial_{m} T_{\brn} 
	+\brPhi_{m \brn}{}^{\brp} T_{\brp}\,.
\ee
Note that covariant derivative $\hcD_{\brm}$ for $\Spin(1,9)$ and $\Spin(9,1+\text{dim}\,G)$ vector representation
\be\ba{ll}
\hcD_{\brm} T_{m} = 
	\partial_{\brm} T_{m} 
	+ \Phi_{\brm n}{}^{p} T_{p}\,,
\qquad 
\hcD_{m} T_{\brn} = \partial_{m} T_{\brn} 
	+\brPhi_{m \brn}{}^{\brp} T_{\brp} 
	+\brPhi_{m \brn}{}^{\bra} T_{\bra}\,,
\\
\hcD_{\bra} T_{m} = \Phi_{\bra n}{}^{p} T_{p}\,.
\ea\ee
 Since the gravitino-curvature $\psi_{[\brm\brn]}$ is a composite field, it does not introduce any fermionic degrees of freedom. For notational convenience, we combine the gauginos and gravitino in $\mathbf{O}(9,1+\text{dim}\,G)$ covariant way as $\psi_{\hat\brp} = \{\psi_{\brm}\,, \psi_{\bra}\}$. 

The Dirac operators for $\Spin(1,9)$ spinors are given by
\be
\gamma^{m} \hcD_{m} \rho\,, ~~~~~~~~\hcD_{\hat\brm} \rho\,, ~~~~~~~~ \gamma^{m} \hcD_{m} \psi_{\hat\brn}\,,
\ee
where the explicit expressions for these are
\be\ba{rl}
\gamma^{m} \hcD_{m} \rho =& \gamma^{m} \partial_{m} \rho 
	+ \tfrac{1}{4} \Phi_{mnp}\gamma^{mnp} \rho 
	+ \half \Phi^{m}{}_{mp} \gamma^{p} \rho\,,
\\
\hcD_{\hbrm} \rho =& \partial_{\hbrm} \rho + \tfrac{1}{4} \Phi_{\hbrm np} \gamma^{np} \rho\,,
\\
\gamma^{m}\hcD_{m} \psi_{\hbrn} = & 
	\gamma^{m} \partial_{m} \psi_{\hbrn} 
	+ \tfrac{1}{4} \Phi_{mnp}\gamma^{mnp} \psi_{\hbrn} 
	+ \half \Phi^{m}{}_{mp} \gamma^{p} \psi_{\hbrn} 
	+ \gamma^{m} \brPhi_{m \hbrn\hbrp} \psi^{\hbrp}\,.
\\ =& \gamma^{m} \partial_{m} \psi_{\hbrn} 
	+ \tfrac{1}{4} \Phi_{mnp}\gamma^{mnp} \psi_{\hbrn} 
	+ \half \Phi^{m}{}_{mp} \gamma^{p} \psi_{\hbrn} 
	+ \gamma^{m} \brPhi_{m \hbrn\brp} \psi^{\brp}
	+ \gamma^{m} \brPhi_{m \hbrn\bra} \psi^{\bra}
	\,.
\ea
\ee

\subsection{SUSY transformation}
We start from SUSY transformation of  $\ODDG$ gauged DFT \cite{Berman:2013cli} 
\be\ba{l}
\delta_{\varepsilon} d = -i\half \brvare \rho\,,
\\
\delta_{\varepsilon} V_{\hM m} = -i \brV_{\hM}{}^{\hbrq} \brvare \gamma_{m} \psi_{\hbrq}\,,
\\
\delta_{\varepsilon} \brV_{\hM\hbrm} = i V_{M}{}^{q} \brvare \gamma_{q} \psi_{\hbrm}\,,
\\
\delta_{\varepsilon} \rho = - \gamma^{m} \hat\cD_{m} \varepsilon\,,
\\
\delta_{\varepsilon} \psi_{\hbrm} = \hat\cD_{\hbrm} \varepsilon\,.
\ea\label{ODDGcovSUSY}\ee
Once we have the double-vielbein parametrization (\ref{parametrizationV}) and (\ref{parametrizationbrV}), it makes sense to decompose $\ODG$ vector indices in  (\ref{ODDGcovSUSY}) as
\be\ba{l}
\delta_{\varepsilon} d = -i\half \brvare \rho\,,
\\
\delta_{\varepsilon} V_{\hM m} = 
	-i \brV_{\hM}{}^{\brq} \brvare \gamma_{m} \psi_{\brq} 
	-i \brV_{\hM}{}^{i} \brvare \gamma_{m} \chi_{i}
	-i \brV_{\hM}{}^{[\brm\brn]} \brvare \gamma_{m} \psi_{\brm\brn}\,,
\\
\delta_{\varepsilon} \brV_{\hM\brm} = i V_{\hM}{}^{q} \brvare \gamma_{q} \psi_{\brm}\,,
\\
\delta_{\varepsilon} \brV_{\hM i} = i V_{\hM}{}^{q} \brvare \gamma_{q} \chi_{i}\,,
\\
\delta_{\varepsilon} \brV_{\hM[\brm\brn]} = i V_{\hM}{}^{q} \brvare \gamma_{q} \psi_{\brm\brn}\,,
\\
\delta_{\varepsilon} \rho = - \gamma^{m} \cD_{m} \varepsilon\,,
\\
\delta_{\varepsilon} \psi_{\brm} = \cD_{\brm} \varepsilon\,,
\\
\delta_{\varepsilon} \chi_{i} = \frac{1}{4}\Phi_{i mn} \gamma^{mn} \varepsilon\,,
\\
\delta_{\varepsilon} \psi_{\brm\brn} =  \frac{1}{4} \Phi_{[\brm\brn] mn} \gamma^{mn} \varepsilon\,.
\ea\label{ODDGcovSUSY2}\ee
Note that this SUSY transformation is identical with the generalized geometry result \cite{Coimbra:2014qaa}. 

However, one can show that (\ref{ODDGcovSUSY2}) is inconsistent with the double-vielbein parametrization as other bosonic symmetries. It should be modified by introducing a compensating local Lorentz transformation.  For instance, $V_{M}{}^{\bra}$ is a derivative index valued vector, but $\delta_{\varepsilon} \brV^{\mu \bra}$ does not vanish. If we take the compensating local Lorentz transformation as
\be
\Lambda_{\brm}{}^{\bra} = 
	i \bar{\varepsilon} \gamma_{m} \psi^{\bra} (\bre_{\brm}{}^{\mu} e_{\mu}{}^{m})\,. 
	\qquad 
\Lambda_{\bra}{}^{\brb} = 0\,.
\label{CompensatingSUSY}\ee
then we have a consistent modified SUSY transformation for $\brV_{M}{}^{\bra}$
\be\ba{rl}
\delta_{\varepsilon} \brV_{M}{}^{\bra} & =
	i V_{M}{}^{m} \bar{\varepsilon}\gamma_{m}\psi^{\bra} 
	+ \brV_{M}{}^{\brm} \Lambda_{\brm}{}^{\bra}
	+ \brV_{M}{}^{\brb} \Lambda_{\brb}{}^{\bra}\,,
\\&
	= i \cV_{M}{}^{m}  \bar{\varepsilon}\gamma_{m}\psi^{\bra} \,,
\ea\label{SUSYVMbra}\ee
where the derivative index valued vector $\cV_{M}{}^{m}$ is defined as
\be
\cV_{M}{}^{m} := V_{M}{}^{m} + \brV_{M}{}^{\brm} (\bre_{\brm}{}^{\mu} e_{\mu}{}^{m}) =
		\sqrt{2} \bpm 0 \\ e_{\mu}{}^{m}\epm\,.
\ee
Furthermore, the compensating local Lorentz transformation (\ref{CompensatingSUSY}) can be applied to the constant component $V_{A}{}^{\bra}$ as well, and the modified SUSY transformation gives a vanishing SUSY variation
\be
\delta_{\varepsilon} \brV_{A}{}^{\bra} 
	= i \big(
	V_{A}{}^{m} + \brV_{A}{}^{\brm} (\bre_{\brm}{}^{\mu} e_{\mu}{}^{m} )
	\big) \bar{\varepsilon} \gamma_{m} \psi^{\bra} = 0\,.
\ee
Therefore the modified SUSY transformation with the local Lorentz transformation (\ref{CompensatingSUSY}) is given by
\be\ba{ll}
\delta_{\varepsilon} d = 
	-i\half \brvare \rho\,,
\\
\delta_{\varepsilon} V_{\hM m} = -i \brV_{\hM}{}^{\brq} \brvare \gamma_{m} \psi_{\brq} 
	-i \brV_{\hM}{}^{i} \brvare \gamma_{m} \chi_{i}
	-i \brV_{\hM}{}^{[\brm\brn]} \brvare \gamma_{m} \psi_{\brm\brn}\,,
\\
\delta_{\varepsilon} \brV_{\hM\brm} = 
	i V_{\hM}{}^{m} \brvare \gamma_{m} \psi_{\brm} 
	+ i\brvare \gamma_{m} \psi_{\bra} (\bre_{\brm}{}^{\mu} e_{\mu}{}^{m}) \brV_{\hM}{}^{\bra}\,,
\\
\delta_{\varepsilon} \brV_{M\bra} = 
	i \cV_{M}{}^{m} \brvare \gamma_{m} \psi_{\bra}\,,
\\
\delta_{\varepsilon} \brV_{A\bra} = 0\,,
\\
\delta_{\varepsilon} \rho = 
	- \gamma^{m} \cD_{m} \varepsilon\,,
\\
\delta_{\varepsilon} \psi_{\brm} = 
	\cD_{\brm} \varepsilon\,,
\\
\delta_{\varepsilon} \psi_{\bra} = 
	\frac{1}{4} \Phi_{\bra mn} \gamma^{mn} \varepsilon\,.
\ea\label{SUSYcovODD}\ee

Now, let's consider how to define the explicit form of the $\cA_{M[\brm\brn]}$. Since it should behave as a gauge field for $\Spin(9,1)$ local Lorentz transformation, the DFT spin-connection $\brPhi_{p\brm\brn}$ would be a good candidate. However there are ambiguities due to a torsion which is not determined yet. 
It is interesting that even if we don't have explicit definition of $\cA_{M[\brm\brn]}$, we can read off $\delta_{\varepsilon} \cA_{M[\brm\brn]}$ from the $\ODD$ structure of double-vielbein (\ref{SUSYcovODD})
\be 
\delta_{\varepsilon} \cA_{\mu [\brm\brn]}  = -i \sqrt{2} (\brvare \gamma_{\mu} \psi_{[\brm\brn]})\,.
\label{SUSYA}\ee
On the other hand the direct computation of $\delta_{\varepsilon} \brPhi_{p\brm\brn}$ gives
\be
\delta_{\varepsilon} \brPhi_{p\brm\brn} = 
	-2i \cD_{[\brm} \big(\brvare \gamma_{|p|} \psi_{\brn]}\big) 
	- i \Phi_{\brq\brm\brn} \brvare \gamma_{p} \psi^{\brq}\,,
\label{SUSYPhi}\ee 
thus we cannot identify $\cA_{p [\brm\brn]}$ and $\brPhi_{p\brm\brn}$. 
To get  a super-covariant transformation as (\ref{SUSYA}), we define $\cA_{M[\brm\brn]}$ by adding gravitinos
\be\ba{ll}
\cA_{\mu[\brm\brn]} & := 
	\sqrt{2} e_{\mu}{}^{p} \big( \brPhi_{p\brm\brn} 
	+ i \brpsi_{\brm} \gamma_{p} \psi_{\brn}\big)\,,
\\&
	= 
	\bromega_{\mu \brm\brn} 
		+ \half H_{\mu\brm\brn} 
		+ i \sqrt{2} \brpsi_{\brm} \gamma_{p} \psi_{\brn}\,,
\ea\label{defA}\ee
then it transform as (\ref{SUSYA}).

Next, we examine the SUSY variation of gravitino curvature $\delta_{\varepsilon} \psi_{[\brm\brn]}$ up to fermion leading order. 
We can read off the $\delta_{\varepsilon} \psi_{[\brm\brn]}$ from (\ref{SUSYcovODD})
\be\ba{ll}
\delta_{\varepsilon} \psi_{[\brm\brn]} &= 
	\tfrac{1}{4} \Phi_{[\brm\brn] mn} \gamma^{mn} \varepsilon\,,
\\ &
	= - \tfrac{1}{8} \sqrt{\alphap} \bar{R}^{H}{}_{mn\brm\brn} \gamma^{mn} \varepsilon\,.
\ea\label{SUSYTransfGraviCurvature1}\ee
On the other hand, the direct computation of $\delta_{\varepsilon} \psi_{[\brm\brn]}$ gives
\be\ba{ll}
\delta_{\varepsilon} \psi_{[\brm\brn]} &= 
	\sqrt{\alphap} \big( [\cD_{\brm}\,,\cD_{\brn}]\varepsilon 
	+ (\Phi^{\brp}{}_{\brm\brn} -\cA^{\brp}{}_{\brm\brn} ) \cD_{\brp} \varepsilon \big)\,,
\\
	&=  \tfrac{1}{8}\sqrt{\alphap} R^{H}{}_{\brm\brn pq} \gamma^{pq} \varepsilon\,,
\ea\label{SUSYTransfGraviCurvature2}\ee
where $R^{H}{}_{\mu\nu mn}$ and $\brR^{H}{}_{\mu\nu \brm\brn}$ are the Riemann tensor with torsion $H$ defined in (\ref{RH}).
Even though (\ref{SUSYTransfGraviCurvature1}) and (\ref{SUSYTransfGraviCurvature2}) are not equivalent, after diagonal gauge fixing, one can show that the difference is nothing but a sub-leading order correction \cite{Bergshoeff:1989de}
\be
R^{\scriptscriptstyle (+)}{}_{\mu\nu\rho\sigma} 
- R^{\scriptscriptstyle (-)}{}_{\rho\sigma\mu\nu} 
	= 2 \partial_{[\mu} H_{\nu\rho\sigma]}\,.
\ee
From the Bianchi identity (\ref{BI}), it is reduced to
\be
R^{\scriptscriptstyle (+)}{}_{\mu\nu\rho\sigma} 
- R^{\scriptscriptstyle (-)}{}_{\rho\sigma\mu\nu} 
	= \alphap T_{\mu\nu\rho\sigma}\,,
\ee
where
\be
T_{\mu\nu\rho\sigma} = \tfrac{3}{4} \Tr\big( F \wedge F - R \wedge R \big)_{[\mu\nu\rho\sigma]}\,.
\ee
Therefore, SUSY transformation of gravitino curvature at order $(\alphap)^{(\frac{3}{2})}$ is given by
\be
\big(\delta \psi_{[\brm\brn]}\big)^{\scriptscriptstyle (\frac{3}{2})} = (\alphap)^{\frac{3}{2}} \tfrac{1}{4} T_{\brm\brn mn} \gamma^{mn} \varepsilon\,.
\label{HigherOrderGravCurvVar}\ee
Or, in other words $\Phi^{\scriptscriptstyle (\frac{3}{2})}{}_{[\brm\brn] mn} = T_{[\brm\brn] mn}$.

\subsection{{SUSY action}}
We may finally turn to the supersymmetric action. Following the supersymmetric gauged DFT \cite{Berman:2013cli}, we have a supersymmetric action which is invariant under (\ref{ODDGcovSUSY})
\be
\cL_{\rm \scriptscriptstyle het} = e^{-2d} \Big[2 S + 4i \big(\brrho \gamma^{m} \hat\cD_{m} \rho -2 \bar{\psi}^{\hbrm} \hat\cD_{\hbrm} \rho -   \brpsi^{\hbrm} \gamma^{m}\hat\cD_{m} \psi_{\hbrm}\big)\Big]\,.
\label{ODDGcovAction}\ee 
The SUSY variation up to leading order in fermions is given by
\be\ba{ll}
\delta \cL_{\rm \scriptscriptstyle het}= e^{-2d} \Big[& 
	- 8 \delta d P^{\hM\hP} P^{\hN\hQ} S_{\hM\hN\hP\hQ} 
	+ 4 \delta P^{\hM\hP} P^{\hN\hQ} S_{\hM\hN\hP\hQ}  
\\&
	+ 8i \brrho \left(  \gamma^{m} \hcD_{m} \delta \rho 
	- \hcD_{\hbrm} \delta\psi^{\hbrm} \right)
	- 8 i \brpsi^{\hbrm} \left(\gamma^{m} \hcD_{m}\delta\psi_{\hbrm}
	+ \hcD_{\hbrm} \delta \rho\right)
\Big]\,.
\ea\ee
Substituting the SUSY transformation in (\ref{ODDGcovSUSY}), one can then show that SUSY invariance is guaranteed from the following identities 
\be\ba{rl}
\gamma^{m}\gamma^{n} \hcD_{m} \hcD_{n} \varepsilon+ \hcD^{\brm} \hcD_{\brm}\varepsilon =& 
	- \tfrac{1}{4}\varepsilon P^{\hM\hN} P^{\hP\hQ} S_{\hM\hQ\hN\hP}\,,
\\
\gamma^{n} \comm{\hcD_{\hbrm}}{\hcD_{n}} \varepsilon =& 
	-\gamma^{n}\varepsilon \brV^{\hM}{}_{\hbrm} V^{\hN}{}_{n} P^{\hP\hQ} S_{\hM\hP\hN\hQ} \,.
\ea\ee

We may rephrase the action (\ref{ODDGcovAction}) and supersymmetry transformation (\ref{SUSYcovODD}) with the explicit parametrization of double-vielbein. For systematic approach, we denote $n$-th order $\alphap$ terms as $(n)$. For instance, the next leading order correction of supersymmetry transformation for gravitino-curvature is of order $\frac{3}{2}$,
\be
\big( \delta \psi_{\brm\brn}\big)^{\scriptscriptstyle (\frac{3}{2})} = (\alphap)^{\scriptscriptstyle (\frac{3}{2})} T_{\brm\brn pq} \gamma^{pq} \varepsilon\,.
\ee 
thus it is denoted by $(\frac{3}{2})$ as a superscription.

We then summarize the Lagrangian and supersymmetry transformations in the leading order correction:
\begin{itemize}
\item Spin-connections
\be\ba{ll}
\Phi^{\scriptscriptstyle (\frac{1}{2})}{}_{i mn} = - \half \sqrt{\alphap} (\cF_{mn})_{i}
\\
\Phi^{\scriptscriptstyle (\frac{1}{2})}{}_{[\brm\brn]mn} = - \half \sqrt{\alphap} \brR^{H}{}_{mn\brm\brn}
\\
\Phi^{\scriptscriptstyle (\frac{3}{2})}{}_{[\brm\brn] mn} = (\alphap)^{\frac{3}{2}} T_{[\brm\brn] mn}\,.
\ea\ee
\item Lagrangian 
\be\ba{cl}
\cL_{B} ^{\scriptscriptstyle (0)}  & = 
	4 \partial^{m} \Phi^{\scriptscriptstyle (0)}{}^{n}{}_{mn} 
	- 2\Phi^{\scriptscriptstyle (0)}{}^{m}{}_{m}{}^{p} \Phi^{\scriptscriptstyle (0)}{}^{n}{}_{np} 
	-3 \Phi^{\scriptscriptstyle (0)}{}^{[mnp]} \Phi^{\scriptscriptstyle (0)}{}_{mnp} 
	- \Phi^{\scriptscriptstyle (0)}{}^{\brp mn} \Phi^{\scriptscriptstyle (0)}{}_{\brp mn} 
\\&\quad
	 - 2 f_{p mn } \Phi^{\scriptscriptstyle (0)}{}^{pmn} 
	 - 2 f_{\brp mn } \Phi^{\scriptscriptstyle (0)}{}^{\brp mn}
	 - 2 f_{\bra mn } \Phi^{\scriptscriptstyle (0)}{}^{\bra mn}
	
\\
\cL_{B} ^{\scriptscriptstyle (1)} & =
	- \Phi^{\scriptscriptstyle (\frac{1}{2})}{}_{i pq} \Phi^{\scriptscriptstyle (\frac{1}{2})}{}^{i pq} 
	+ \Phi^{\scriptscriptstyle (\frac{1}{2})}{}_{[\brm\brn] pq} \Phi^{\scriptscriptstyle (\frac{1}{2})}{}^{[\brm\brn] pq}
\\
\cL_{F}^{\scriptscriptstyle (0)} & = 
	4i \brrho \gamma^m \cD_{m} \rho 
	- 8 i \brpsi^{\brm} \cD_{\brm} \rho 
	-4i \brpsi^{\brm} \gamma^{m} \cD_{m} \psi_{\brm}\,,
\\
\cL_{F}^{\scriptscriptstyle (1)} & = 
	-2i \brchi^{i} \gamma^{pq} \rho\Phi^{\scriptscriptstyle (\frac{1}{2})}{}_{i pq}	
	+2i \brpsi^{[\brm\brn]} \gamma^{pq} \rho\Phi^{\scriptscriptstyle (\frac{1}{2})}{}_{[\brm\brn] pq}	
	-4i \brchi^{i} \gamma^m \cD_{m} \chi{i} 
\\&\quad

	+4i \brpsi^{[\brm\brn]} \gamma^m \cD_{m} \psi_{[\brm\brn]} 
	-4i \brpsi^{\brm}\gamma^{m}\chi^{i}  \brPhi^{\scriptscriptstyle (\frac{1}{2})}{}_{m\brm i}
	+8i \brpsi^{\brm}\gamma^{m}\psi^{[\brm\brn]}  \brPhi^{\scriptscriptstyle (\frac{1}{2})}{}_{m\brm[\brm\brn]}\,.
\ea\label{LeadingAction}\ee
\item Supersymmetry transformations for fermion fields
\be\ba{rl}
\big(\delta \rho\big)^{\scriptscriptstyle (0)} 
	&= - \gamma^{m} \cD_{m} \varepsilon\,,
\\
\big(\delta \psi_{\brm} \big)^{\scriptscriptstyle (0)}  
	&= \cD_{\brm} \varepsilon\,,
\\
\big( \delta\chi_{i} \big)^{\scriptscriptstyle (\frac{1}{2})}
	&= \tfrac{1}{4} \Phi^{\scriptscriptstyle (\frac{1}{2})}{}_{i pq} \gamma^{pq} \varepsilon\,,
\\
\big( \delta\psi_{[\brm\brn]} \big)^{\scriptscriptstyle (\frac{1}{2})}
	&= \tfrac{1}{4} \Phi^{\scriptscriptstyle (\frac{1}{2})}{}_{[\brm\brn] pq} \gamma^{pq} \varepsilon\,,
\ea\label{leadingSUSYtransf}\ee
\end{itemize}
where the covariant derivative $\cD_{\brm}$ and $\cD_{m}$ gives zeroth order contribution.
Here we have ignored all intrinsic $\alphap$-contributions which are located in the $H$ fields, because the explicit form of $H$ is not relevant in the fermion leading order calculation. 

From the fact that the SUSY transformation of gravitino-curvature receives further correction in the order of $(\alphap)^{\frac{3}{2}}$,  the next order correction in $\delta \cL$ arises naturally, and we discuss this point in the next section.

\section{Quadratic $\alphap$-corrections}
In the previous section we have shown that the SUSY variation of the gravitino curvature receives higher order contributions, $(\delta\psi_{[\brm\brn]})^{\order{\frac{3}{2}}}$, in (\ref{HigherOrderGravCurvVar}). It leads to further $(\alphap)^{2}$-order corrections on the SUSY variation of $\cL_F^{\order{1}}$
\be\ba{ll}
\big(\delta \cL_{F}^{\order{1}}\big)^{\scriptscriptstyle (2)} 
=(\alphap)^2 \Big[ & 
	i\half \brrho \gamma^{mn}\gamma^{pq} \varepsilon \Phi^{\scriptscriptstyle (\frac{1}{2})}{}_{\brm\brn mn} T^{\brm\brn}{}_{pq}  
	-2i \cD_{m} \brpsi^{\brm\brn} \gamma^{m} \gamma^{pq} \varepsilon T_{\brm\brn pq} 
\\&
	+2i \brpsi^{\brm} \gamma^{m} \gamma^{pq} \varepsilon \brPhi^{\scriptscriptstyle (\frac{1}{2})}{}_{m\brm[\brn\brp]} T^{\brn\brp}{}_{pq}\Big]\,.
\ea\label{residue1}\ee
In order to cancel out this additional variation, we should introduce higher order $\alpha'$-corrections in the SUSY transformation and the corresponding action. In keeping with the fact that the bosonic Lagrangian does not receive $R^{3}$ correction at order $(\alphap)^2$ \cite{Bergshoeff:1989de}, we will consider the fermionic sector first. 

Let us assume that the supersymmetry transformation of the gravitino $\psi_{\brm}$ is given by the following general form:
\be
(\delta \psi_{\brm})^{\scriptscriptstyle (2)} 
	=\tfrac{1}{4}  \Phi^{\scriptscriptstyle (2)}{}_{\brm mn}\gamma^{mn}\varepsilon\,,
\label{gravitinoSUSY(2)}\ee
where $\Phi^{\scriptscriptstyle (2)}{}_{\brm mn}$ should be determined from the closure of SUSY. If we substitute (\ref{gravitinoSUSY(2)}) into the $\cL_F^{\scriptscriptstyle (0)}$, we have 
\be\ba{ll}
\big(\delta \cL_{F}^{\scriptscriptstyle (0)}\big)^{\scriptscriptstyle (2)} 
	= (\alphap)^2 e^{-2d}\Big[  
		2i \cD_{\brm} \brrho \gamma^{mn} \varepsilon \Phi^{\scriptscriptstyle (2)}{}^{\brm}{}_{mn}
		+ 2i \cD_{m} \brpsi^{\brm} \gamma^{m} \gamma^{pq} \varepsilon \Phi^{\scriptscriptstyle (2)}{}_{\brm pq}	
	\Big]\,,
\ea\label{alpha2SF0}\ee
and one can show that it is insufficient to cancel (\ref{residue1}).  

Next, we introduce the following correction to the fermion Lagrangian
\be
\cL_{F}^{\scriptscriptstyle (2)} 
	= (\alphap)^2 e^{-2d}\Big[  
		4i\brpsi^{\brm} \gamma^{pq} \big(\cD^{\brn} \rho + \gamma^{r} \cD_{r} \psi^{\brn} \big) T_{\brm\brn pq}
	\Big]\,,
\label{ActionCorrection}\ee
which is proportional to the equation of motion of $\psi_{\brm}$ at order $(\alphap)^0$ \cite{Bergshoeff:1989de}. Then $\cL_{F}^{\scriptscriptstyle (2)}$ transform under  (\ref{leadingSUSYtransf}) at order $(\alphap)^2$ as
\begin{multline}
\delta^{\scriptscriptstyle (0)} \cS^{\order{2}}_{F} = 
	(\alphap)^2 \int 4i e^{-2d}\Big[
		-\brvare\gamma^{np} \big(\cD^{\brm}\cD^{\brn}\rho 
		+ \cD^{\brm} \gamma^{q} \cD_{q} \psi^{\brn} \big)
			T_{\brm\brn np}
\\
		- \brvare\gamma^{np} \big( \cD^{\brn} \rho 
		+ \gamma^{q} \cD_{q} \psi^{\brn} \big) \cD^{\brm} 
			T_{\brm\brn np}
		+\brpsi^{\brm} \gamma^{np} \gamma^{q} \varepsilon S^{\order{0}}{}^{\brn r}{}_{q r} 
		T_{\brm\brn np}
		\Big]\,.
\label{delta0SF2-1}\end{multline}
By using integration by parts and the following identities
\be\ba{ll}
\comm{\cD_{\brm}}{\cD_{\brn}} \varepsilon =
	- \brPhi^{\order{0}}{}^{\brp}{}_{\brm\brn} \cD_{\brp}\varepsilon 
	+ \cA^{\brp}{}_{[\brm\brn]} \cD_{\brp} \varepsilon 
	+ \tfrac{1}{8} R^{H}{}_{\brm\brn pq} \gamma^{pq}\varepsilon\,,
\\
\comm{\cD^{\brm}}{\cD_{p}} \psi^{\brn} = 
	2 S^{\brm}{}_{p}{}^{\brn}{}_{\brq} \psi^{\brq}
	+ \half S^{\brm}{}_{pqr} \gamma^{qr} \psi^{\brn}\,,
\ea\ee
the SUSY variation (\ref{delta0SF2-1}) can be rewritten as
\begin{equation}
\ba{ll}
\delta^{\scriptscriptstyle (0)} \cS_{F}^{\scriptscriptstyle (2)} =
	(\alphap)^2 \displaystyle\int 4i e^{-2d}\Big[&
		-\half \brvare\gamma^{mn} \big(\tfrac{1}{8} R^{H}{}_{\brm\brn pq} \gamma^{pq} \rho 
		-X^{\brp}{}_{\brm\brn} \cD_{\brp} \rho \big) T^{\brm\brn}{}_{mn}
\\&
	+ \brvare \gamma^{np} \gamma^{q} \cD_{q} 
	\big(\half \psi^{\brm\brn}  
	+\half X^{\brp\brm\brn} \psi_{\brp}\big) 	T_{\brm\brn np}	
\\&
	-2\brvare \gamma^{np}\gamma^{q} \psi^{\brq} S^{\brm}{}_{q}{}^{\brn}{}_{\brq} T_{\brm\brn np}	
	+ \brvare \gamma^{np} \gamma^{q} \brpsi^{\brn} S^{\brm r}{}_{q r} T_{\brm\brn np}	
\\&
	- \brvare\gamma^{np} \big( \cD^{\brn} \rho 
	+ \gamma^{q} \cD_{q} \psi^{\brn} \big) 
		\cD^{\brm} T_{\brm\brn np}		
	+\brpsi_{\brm} \gamma^{np} \gamma^{q} \varepsilon S^{\brn r}{}_{q r} 
	T_{\brm\brn np}	 \Big]\,,
\ea\label{delta0SF2-2}\end{equation}
where 
\be
X^{\order{0}}{}^{\brp}{}_{\brm\brn}  = \brPhi^{\order{0}}{}^{\brp}{}_{\brm\brn}-\cA^{\brp}{}_{[\brm\brn]} \,.
\ee

Next, if we define $\Phi^{\scriptscriptstyle (2)}{}_{\brp mn}$ as
\be
\Phi^{\scriptscriptstyle (2)}{}_{\brm pq} := \half \big(
	\cD^{\brn} T_{\brn\brm pq} 
	-\half X_{\brm}{}^{\brp\brq} T_{\brp\brq pq}\big) \,.
\ee
and use the following identity
\be
\cD_{q} X^{\brp\brm\brn}  =
	  2 S^{\order{0}}{}_{q}{}^{\brp\brm\brn} 
	+ \half \brR^{H}{}_{q}{}^{\brp\brm\brn} \,.
\ee
then the SUSY variation (\ref{delta0SF2-2}) reduces to
\be\ba{ll}
\delta^{\scriptscriptstyle (0)} \cS_{F}^{\scriptscriptstyle (2)} =
	(\alphap)^2 \displaystyle\int 2i e^{-2d}\Big[&
		-\tfrac{1}{16} \brvare\gamma^{mn} \gamma^{pq} \rho R^{H}{}_{\brm\brn pq} T^{\brm\brn}{}_{mn}
		- \brvare \gamma^{np} \cD^{\brm} \rho \Phi^{\scriptscriptstyle (2)}{}_{\brm np}
\\&
	- \half \brvare \gamma^{np} \gamma^{q} \cD_{q} \psi^{\brm\brn} T_{\brm\brn np}
	+\half \brvare \gamma^{np} \gamma^{q} \psi_{\brp} \cD_{q} X^{\brp\brm\brn} T_{\brm\brn np}
\\&
	- 2 \brvare \gamma^{np}\gamma^{q} \psi^{\brq} S^{\brm}{}_{q}{}^{\brn}{}_{\brq} T_{\brm\brn np}	
	+ 4 \brvare \gamma^{n} \brpsi^{\brn} S^{\brm r p}{}_{r} T_{\brm\brn np}	
\\&
	- \brvare\gamma^{np} \gamma^{q} \cD_{q} \psi^{\brn} \Phi^{\scriptscriptstyle (2)}{}_{\brn np}
	 \Big]\,.
\ea\label{delta0SF21}\ee

Finally, by combining all the terms (\ref{residue1}), (\ref{alpha2SF0}) and (\ref{delta0SF21}) together, we get a simple result
\be
\big(\delta \cS_{F} \big)^{\scriptscriptstyle (2)} = 
	(\alphap)^2 \displaystyle\int 16 i e^{-2d} \brvare \gamma^{q} \psi^{\brn} S^{\scriptscriptstyle (0)}{}^{\brm r p}{}_{r}T_{\brm\brn pq}\,.
\label{residual2}\ee
Recall that an arbitrary variation of the generalized curvature scalar $S$ is proportional to the generalized Ricci tensor $S_{MN}$. Since the variation (\ref{residual2}) is proportional to the generalized Ricci tensor, one can speculate that it is canceled out by introducing $(\alpha')^2$-order corrections to the SUSY variation of the double-vielbein $(\delta V^{\order{0}})^{\order{2}}$, where $\brV^{\order{0}}{}_{M}{}^{\brm}$ is the double-vielbein for ungauged DFT without gauge connections
\be
\brV^{\order{0}}{}_{M}{}^{\brm} = 
	\tfrac{1}{\sqrt{2}} \begin{pmatrix} 
		(\bre^{-1})^{\mu \brm} \\ 
		\bre_\mu{}^{\brm} + B_{\mu\nu} (\bre^{-1})^{\nu \brm}  
	\end{pmatrix}\,.
\ee
Under an arbitrary variation $(\delta V^{\order{0}})^{\order{2}}$, the bosonic action $\cS^{\order{0}}_B$ changes by
\be
\big(\delta\cS^{\scriptscriptstyle (0)}_{B}\big)^{\order{2}}  = 
	\displaystyle\int 8\big(
		\delta V^{\order{0}}{}_{M}{}^{m}\big)^{\scriptscriptstyle(2)} V^{\order{0}}{}^{N}{}_{m} S^{\scriptscriptstyle(0)}{}^{M}{}_{N}\,,
\ee
and if we choose $\big(\delta V^{\order{0}}{}_{M m}\big)^{\scriptscriptstyle(2)}$ as
\be
\big(\delta_{\varepsilon} V^{\order{0}}{}_{M m}\big)^{\scriptscriptstyle (2)} = 
	- 2i (\alphap)^2 \brvare \gamma^{n} \psi^{\brn} \brV^{\order{0}}{}_{\hM}{}^{\brm} T_{\brm\brn mn}\,,
\label{VarcorrV}\ee
then the SUSY variation of the bosonic action at order $(\alphap)^2$ is given by
\be
\big(\delta_{\varepsilon} \cS^{\scriptscriptstyle (0)}_{B} \big)^{\scriptscriptstyle(2)} 
	= -16i (\alphap)^2 \displaystyle\int  \brvare \gamma^{n} \psi^{\brn} T_{\brm\brn mn} S^{\scriptscriptstyle(0)}{}^{\brm m}\,.
\ee
Therefore the $\big(\delta\cS^{\scriptscriptstyle (0)}_{B}\big)^{\order{2}}$ exactly cancels out the $(\delta \cS_F^{\order{0}})^{\order{2}}$ in (\ref{residual2}), and it shows that supersymmetry is closed at $(\alphap)^2$-order. 

However, we need to determine the missing SUSY transformations at $(\alpha')^2$-order. For example, we can obtain $\big(\delta_{\varepsilon} \brV^{\order{0}}{}_{M}{}^{\brm}\big)^{\order{2}}$ by requiring that SUSY variations of the component fields from the $\big(\delta_{\varepsilon} \brV^{\order{0}}{}_{M}{}^{\brm}\big)^{\order{2}}$ and $\big(\delta V^{\order{0}}{}_{M m}\big)^{\scriptscriptstyle(2)}$ should be equivalent, as a consistency condition, such as
\be\ba{cl}
\big(\delta_{\varepsilon} d \big)^{\scriptscriptstyle(2)} &= 0\,,
\\
\big(\delta_{\varepsilon} e_{\mu m} \big)^{\scriptscriptstyle(2)} &= 
	2i (\alphap)^{2} \brvare \gamma^{n} \psi^{\brn} T_{\mu\brn mn}\,,
\\
\big(\delta_{\varepsilon} B_{\mu \nu} \big)^{\scriptscriptstyle(2)} &= 
	2i (\alphap)^{2} \brvare \gamma^{n} \psi^{\brn} T_{\mu\nu \brn n}\,.
\ea\ee
From this condition, $\big(\delta_{\varepsilon} \brV^{\order{0}}{}_{M}{}^{\brm}\big)^{\order{2}}$ is given by
\be
\big(\delta_{\varepsilon} \brV^{\order{0}}{}_{M\brm}\big)^{\order{2}} = 
	2i \brvare \gamma^{n} \psi^{\brn} V^{\order{0}}{}_{M}{}^{m} T_{\brm\brn mn}\,.
\ee
However, it is still not sufficient to determine the other components, $\big(\delta_{\varepsilon} V_{A}{}^{m}\big)^{\order{2}}$, $\big(\delta_{\varepsilon} \brV_{M}{}^{\bra}\big)^{\order{2}}$ and $\big(\delta_{\varepsilon} \brV_{A}{}^{\brm}\big)^{\order{2}}$. To determine these we should consider $(\alphap)^3$-corrections, thus we will leave this issue for future work. 

In summary, the quadratic $\alpha'$-correction of the action is
\be
\cL_{F}^{\scriptscriptstyle (2)} 
	= (\alphap)^2 2i e^{-2d}\Big[  
		\brpsi^{\brm} \gamma^{pq} \big(\cD^{\brn} \rho + \gamma^{r} \cD_{r} \psi^{\brn} \big) T_{\brm\brn pq}
	\Big]\,,
\label{ActionCorrection}\ee
and the SUSY variations are
\be\ba{cl}
\big(\delta_{\varepsilon} V_{M m}\big)^{\scriptscriptstyle (2)} & =
	- 2i \brvare \gamma^{n} \psi^{\brn} \brV_{M}{}^{\brm} T_{\brm\brn mn}\,,
\\
\big( \delta_{\varepsilon} \brV_{M\brm}\big)^{\scriptscriptstyle (2)} &= 
	2i \brvare \gamma^{n} \psi^{\brn} V_{M}{}^{m} T_{\brm\brn mn}\,,
\\
\big( \delta_{\varepsilon} \psi_{\brm} \big)^{\scriptscriptstyle (2)}  &= 
	\frac{1}{8} \big(\cD^{\brn} T_{\brn\brm pq} 
	-\half X_{\brm}{}^{\brp\brq} T_{\brp\brq pq}\big) \gamma^{pq} \varepsilon
\\
\big( \delta_{\varepsilon} \psi_{[\brm\brn]} \big)^{\scriptscriptstyle (\frac{3}{2})}  & =
	\frac{1}{4} T_{\brp\brq pq} \gamma^{pq} \varepsilon\,.
\ea\label{SUSYcovODD2}\ee
\section{Conclusions}

In this paper we have shown that the $\mathbf{O}(1,D-1)\times\ODG$ double-vielbein formalism for gauged DFT can be applied to the extended tangent space formalism for leading order $\alpha'$-corrections in heterotic DFT. Associated spin-connections, the generalized Ricci tensor and the generalized Ricci scalar have been introduced, and using this framework we have constructed supersymmetric heterotic DFT with leading $\alpha'$-corrections. By solving the defining properties of the double-vielbein, we got a canonical parametrization in terms of physical component fields, $e^{a},~B,~A,$ and $\brPhi^{[\brm\brn]}$. The $R^2$ term arises as a field strength of the $\mathbf{O}(1,9)$ gauge field, $\brPhi^{[\brm\brn]}$ only after the parametrization. Thus in order to describe the leading $\alpha'$-corrections, the parametrization is essential. Under the symmetries of heterotic DFT such as $\ODDG$, the generalized Lie derivative and supersymmetry the parametrization is not maintained. We have obtained explicit modifications of the symmetry transformations to be compatible with the parametrization. 

We have shown that the extended tangent space formalism can be extended to the next order $\alphap$-correction. We have checked that there is the hidden $(\alpha')^{\frac{3}{2}}$-correction in the supersymmetry transform, $(\delta_{\varepsilon} \psi_{\brm\brn})^{\order{\frac{3}{2}}}$. Thus additional $(\alpha')^2$ terms arise in the SUSY variation, and we have found  the corrections in the action and SUSY transformation which cancel out the terms. As a result, we have shown that there is no $R^3$ correction in the bosonic part, and the fermionic sector is proportional to the equations of motion for the gravitino at order $(\alphap)^0$. This is consistent with the heterotic supergravity result \cite{Bergshoeff:1989de}.

We can continue the SUSY closure process, and it may suggest that the gauged DFT description is valid even in higher order $\alpha'$-corrections. However, the full $(\alpha')^3$-corrections consist of two parts \cite{Tseytlin:1995bi}: anomaly-related terms containing $(\tr F^2 - \tr R^2)$ and without Yang-Mills counterpart term. In the extended tangent space formalism, the $R^2$ term is always involved with $\tr F^2$, thus the $\ODDG$ gauged DFT description makes sense only for anomaly-related terms. Thus it is not obvious how to construct the other sector. It may require a totally different formalism other than the gauged DFT description. 

It is also interesting to study $\alpha'$-corrections of type II DFT. The extended tangent space formalism suggests $\mathbf{O}(10+\text{dim} G_1, 10+\text{dim} G_2)$ gauged DFT, where $G_1$ and $G_2$ are $\mathbf{O}(1,9)$ and $\mathbf{O}(9,1)$ local Lorentz groups respecively. Then the local structure group is given by $\mathbf{O}(1+ \text{dim} \,G_1,9)\times\mathbf{O}(9 ,1+ \text{dim} \,G_2) $ as a maximal compact subgroup. However, it has been shown that this generalized geometry does not admit any consistent torsion free connection \cite{Coimbra:2014qaa}. Furthermore, there is no natural way to define a spinor and Clifford algebra unlike in the heterotic DFT. Since heterotic and type II supergravities share common $\alpha'$-corrections which do not include anomaly-related terms \cite{Tseytlin:1995bi}, $\alpha'$-corrections for the heterotic DFT may give some clues for the type II case.

As we have seen, supergravities are strongly restricted by the $\Odd$ structure even if we consider $\alpha'$-corrections. Usually supersymmetric $\alpha'$-corrections are constructed by supersymmetry completion for a given ansatz which contains all possible terms with arbitrary coefficients. Since the $\Odd$ structure provides a further constraint, the ansatz is greatly simplified. Therefore, it may be possible to construct much higher order $\alphap$-corrections and find a deeper structure of the general $\alpha'$-corrections through the supersymmetry completion method in DFT.

\bigskip
\bigskip

\noindent{{\textbf{Acknowledgements.}}  

I would like to thank to Imtak Jeon and Jeong-Hyuck Park for useful discussions and comments.

\appendix
\section{Conventions \label{appendixA}}
In this section we describe various conventions and indices in detail. We decompose $\ODDG$ covariant quantities to $\ODD$ subgroup.
Here all the hatted indices represent $\ODDG$ covariant quantities:
\begin{enumerate}
\item $\ODDG$ indices :
\begin{itemize}
	\item $\hM,\hN, \cdots$ : $\ODDG$ vector indices,
	\item $ m, n, \cdots$ : Local $\OoD$ vector indices,
	\item $\hbrm,\hbrn, \cdots$ : Local $\ODG$ vector indices.
\end{itemize}
\item After explicit breaking of $\ODDG$ into $\ODD$:
\begin{itemize}
	\item $M,N,\cdots $ : $\ODD$ vector indices,
	\item $m, n, \cdots$ : Local $\OoD$ vector indices,
	\item $\brm,\brn, \cdots$ : Local $\mathbf{O}(D-1,1)$ vector indices.
	\item $\bra,\brb, \cdots$ : Adjoint indices for gauge group $G$.
\end{itemize}
\end{enumerate}
Since the gauge group $G$ is given by the product of two groups $G = G_1 \times G_2$, the gauge indices should be decomposed.
Here $k_{\alpha\beta} = (t^{i})_{\alpha} (t_i)_{\beta}$ and $\tilde{k}_{\tilde\alpha \tilde\beta} = (\tilde{t}^{[\bar{m} \bar{n}]})_{\tilde\alpha} (\tilde{t}_{[\brm\brn]})_{\tilde{\beta}} $ are killing metrics for each gauge group, and $(t^{i})_{\alpha}$ and $(\tilde{t}^{[\brm\brn]})_{\tilde\alpha}$ are the structure constants of the gauge group $G_1$ and $G_2$ respectively
\be
(t^{i})_{kl} = f_{kl}{}^{i} \,, \qquad (t^{[\brm\brn]})_{[\brp\brq][\brr\brs]} = f^{[\brm\brn]}{}_{[\brp\brq][\brr\brs]}
\ee
Thus $t_{i}{}^{\alpha}$ and $t_{[\brm\brn]}{}^{\tilde{\alpha}}$ satisfy
\be
	\comm{t_i}{t_j} = f_{ij}{}^{k} t_{k} \,, 
	\qquad 
	\comm{t_{[\brm\brn]}}{ t_{[\brp\brq]}} = f_{[\brm\brn][\brp\brq]}{}^{[\brr\brs]} t_{[\brr\brs]}
\label{LieAlgebra}\ee
Since the $\OoD$ local Lorentz transformation acts as
\be
\delta_{\Lambda} V_{M [\brm\brm']} = 
	- f_{[\brm\brm'] [\brn\brn']}{}^{[\brp\brp']} \brV_{M}{}^{[\brn\brn']} \Lambda_{[\brp\brp']}  = 
		\Lambda_{\brm}{}^{\brq}\brV_{M[\brq\brm']} + \Lambda_{\brm'}{}^{\brq} \brV_{M[\brm\brq]}
\ee
where we have assume that the pair of indicis $\brm$ and $\brm'$ are antisymmetric to each other, namely $F_{\brm\brm'}= \half (F_{\brm\brm'} - F_{\brm'\brm})$ for an arbitrary $F_{\brm\brm'}$.
Then the structure constant $f_{[\brm\brn][\brp\brq]}{}^{[\brr\brs]}$ is given by
\be
f_{[\brm\brm'] [\brn\brn']}{}^{[\brp\brp']} =  
	- \delta_{\brn}{}^{\brp} \breta_{\brn'\brm} \delta_{\brm'}{}^{\brp'} 
	- \delta_{\brn}{}^{\brp'} \breta_{\brn'\brm'} \delta_{\brm}{}^{\brp}
		= - 2 \delta_{\brn}{}^{\brp} \breta_{\brn'\brm} \delta_{\brm'}{}^{\brp'}
\label{StructureConstant}\ee
One can show that (\ref{StructureConstant}) satisfies Jacobi identity
\be
f_{[\brm\brm'] [\brp\brp']}{}^{[\brq\brq']} f_{[\brn\brn'] [\brq\brq']}{}^{[\brr\brr']} 
+ f_{[\brp\brp'] [\brn\brn']}{}^{[\brq\brq']} f_{[\brm\brm'] [\brq\brq']}{}^{[\brr\brr']}
+ f_{[\brn\brn'] [\brm\brm']}{}^{[\brq\brq']} f_{[\brp\brp'] [\brq\brq']}{}^{[\brr\brr']}
= 0\,,
\ee
thus the $\OoD$ algebra (\ref{LieAlgebra}) is also satisfied.


\section{Supergravity representation}

The $\Spin(1,9)$ Clifford algebra,
\be
(\gamma^{m})^{*} = \gamma^{m} \,, 
\qquad 
\gamma^{m} \gamma^{n} + \gamma^{n} \gamma^{m} = 2 \eta^{mn}\,,
\ee
and chirality operator $\gamma^{(11)} = \gamma^{0} \gamma^{1} \cdots \gamma^{9}$. The symmetric charge conjugation matrice, $C_{\alpha\beta} = C_{\beta\alpha}$, meets 
\be
(C\gamma^{p_{1}p_{2}\cdots p_{n}})_{\alpha\beta}=(-1)^{n(n-1)/2}(C\gamma^{p_{1}p_{2}\cdots p_{n}})_{\beta\alpha}\,,
\ee
and define the charge-conjugated spinors,
\be
\brpsi_{\brp\alpha}=\psi_{\brp}^{\,\beta}C_{\beta\alpha}\,,\qquad 
\brpsi_{\bra\alpha}=\psi_{\bra}^{\,\beta}C_{\beta\alpha}\,,\qquad
\brrho_{\alpha}=\rho^{\beta} C_{\beta\alpha}\,.
\ee
The gravitino and dilatino are set to be Majorana-Weyl spinors of the fixed chirality,
\be
\gamma^{(11)} \psi_{\brp} = \psi_{\brp} \,, \qquad 
\gamma^{(11)} \psi_{\bra} = \psi_{\bra} \,, \qquad 
\gamma^{(11)} \rho = -\rho\,, \qquad
\gamma^{(11)} \varepsilon = \varepsilon
\ee
Using the parametrization of the double-vielbein,  the previous physical spin-connections (\ref{covPhi}) are rewritten in terms of supergravity fields
\be\ba{cl}
\Phi_{[mnp]} &= \frac{1}{\sqrt{2}}\big( \omega_{[mnp]} + \frac{1}{6} H_{mnp} \big)\,,
\\
\Phi_{\brp mn} &=  \frac{1}{\sqrt{2}}\big( \omega_{\brp mn} + \frac{1}{2} H_{\brp mn} \big)\,,
\\
\Phi_{i mn} &=  - \frac{1}{2} \sqrt{\alphap} (F_{mn})_{i}\,,
\\
\Phi_{[\brm\brn] mn} &=  - \frac{1}{2} \sqrt{\alphap} \brR^{H}{}_{mn \brm\brn}\,,
\\
\brPhi_{[\brm\brn\brp]} &= \frac{1}{\sqrt{2}}\big( \bromega_{[\brm\brn\brp]} + \frac{1}{6} H_{\brm\brn\brp} \big)\,,
\\
\brPhi_{[\brm\brn i]}& = -  \frac{1}{6} \sqrt{\alphap}  (F_{\brm\brn})_{i}\,,
\\
\brPhi_{[\brm\brn [\brp\brq]]}& = - \frac{1}{6} \sqrt{\alphap}  \brR^{H}{}_{\brm\brn\brp\brq}\,,
\\
\brPhi_{[\brm[\brp\brp'][\brq\brq']]}& = f_{[\brp\brp'][\brq\brq'][\brr\brr']} \cA_{\brm}{}^{[\brr\brr']}\,,
\\
\brPhi_{[\brm ij]}& = f_{ijk}\cA_{\brm}{}^{k}\,,
\\
\brPhi_{[ijk]}& = \frac{1}{\sqrt{\alphap}} f_{ijk}\,,
\\
\brPhi_{[[\brp\brp'][\brq\brq'][\brr\brr']]}& = \frac{1}{\sqrt{\alphap}} f_{[\brp\brp'][\brq\brq'][\brr\brr']}\,,
\\
\brPhi_{p \brm\brn} &=  \frac{1}{\sqrt{2}}\big( \bromega_{p \brm\brn} + \frac{1}{2} H_{p \brm\brn} \big)\,,
\\
\brPhi_{m\brn i} & = -\half  \sqrt{\alphap}(F_{m\brn})_{i}
\\
\brPhi_{m\brn [\brp\brq]} & = -\half  \sqrt{\alphap} \brR^{H}{}_{m\brn \brp\brq}
\\
\brPhi_{mij} & = \cA_{m}{}^{k} f_{ijk}\,,
\\
\brPhi_{m[\brp\brp'][\brq\brq']} & = \cA_{m}{}^{[\brr\brr']} f_{[\brp\brp'][\brq\brq'][\brr\brr']}\,,
\ea\label{spinconnections}\ee 
where  $H_{mnp}$ is that
\be
H_{\mu\nu\rho} = 3\partial_{[\mu} B_{\nu\rho]} - \alphap \Omega_{\mu\nu\rho}
\ee
and $\Omega_{\mu\nu\rho}$ is the Chern-Simons 3-form which is defined
\be
\Omega_{\mu\nu\rho} = 3\big(\cA_{[\mu}{}^{\bra}\partial_{\nu} \cA_{\rho] \bra} -\tfrac{1}{3} \cA_{\mu}{}^{\bra} \cA_{\nu}{}^{\brb} \cA_{\rho}{}^{\brc} f_{\bra\brb\brc} \big)\,.
\label{CS}\ee

Here $\omega_{\mu mn}$ and $\bromega_{\mu \brm\brn}$ are spin-connections for Riemannian geometry with respect to $e_{\mu}^{m}$ and $\bre_{\mu}^{\brm}$ respectively, and $R^{H}{}_{\mu\nu mn}$ and $\brR^{H}{}_{\mu\nu\brm\brn}$ are Riemann tensors with torsionful connections
\be\ba{ll}
R^{H}{}_{\mu\nu mn} =
	\partial_{\mu} \omega^{H}{}_{\nu mn}
	- \partial_{\nu} \omega^{H}{}_{\mu mn}
	+ \omega^{H}{}_{\mu n}{}^{p} \omega^{H}{}_{\nu pn}
	- \omega^{H}{}_{\nu n}{}^{p} \omega^{H}{}_{\mu pn}
\\
\brR^{H}{}_{\mu\nu \brm\brn} = 
	\partial_{\mu} \bromega^{H}{}_{\nu \brm\brn}
	- \partial_{\nu} \bromega^{H}{}_{\mu \brm\brn}
	+ \bromega^{H}{}_{\mu \brn}{}^{\brp} \bromega^{H}{}_{\nu \brp \brn}
	- \bromega^{H}{}_{\nu \brn}{}^{\brp} \bromega^{H}{}_{\mu \brp \brn}
\ea\label{RH}\ee
where $\omega^{H}{}_{\mu mn}$ and $\bromega^{H}{}_{\mu \brm\brn}$
\be
\omega^{H}{}_{\mu mn} = \omega_{\mu mn} + \half H_{\mu mn}\,, 
\qquad 
\bromega^{H}{}_{\mu \brm\brn} = \bromega_{\mu \brm\brn} + \half H_{\mu \brm\brn}\,.
\ee
After diagonal gauge fixing the torsionful spin-connections $\omega^{H}{}_{\mu}$ are reduced
\be\ba{ll}
\omega^{H}{}_{\mu}{}^{m}{}_{n} \rightarrow &
	\omega^{\scriptscriptstyle (+)}{}_{\mu}{}^{m}{}_{n} 
		=  \omega_{\mu}{}^{m}{}_{n} 
		+ \half H_{\mu}{}^{m}{}_{n}\,,
\\
\bromega^{H}{}_{\mu}{}^{\brm}{}_{\brn} \rightarrow &
 	\omega^{\scriptscriptstyle (-)}{}_{\mu}{}^{m}{}_{n} 
		=  \omega_{\mu}{}^{m}{}_{n} 
		- \half H_{\mu}{}^{m}{}_{n}\,,
\\
\bromega^{H}{}_{\mu \brm\brn} \rightarrow & 
	 - \omega^{\scriptscriptstyle (-)}{}_{\mu mn} 
		=  -\omega_{\mu mn} 
		+ \half H_{\mu mn}\,,

\ea\ee
and the curvature tensors in (\ref{RH}) are reduced to
\be
R^{H}{}_{\mu\nu mn} \rightarrow R^{(+)}{}_{\mu\nu mn} \,, 
\qquad 
\brR^{H}{}_{\mu\nu \brm\brn} \rightarrow -  R^{(-)}{}_{\mu\nu mn}\,.
\ee
where
\be
R^{(\pm)}{}_{\mu\nu}{}^{m}{}_{n} = \partial_{\mu} \omega^{(\pm)}{}_{\nu}{}^{m}{}_{n}
	- \partial_{\nu} \omega^{(\pm)}{}_{\mu}{}^{m}{}_{n}
	+ \omega^{(\pm)}{}_{\mu}{}^{m}{}_{p} \omega^{(\pm)}{}_{\nu}{}^{p}{}_{n}
	-  \omega^{(\pm)}{}_{\nu}{}^{m}{}_{p} \omega^{(\pm)}{}_{\mu}{}^{p}{}_{n}
\ee
Then generalized curvature scalar and generalized Ricci tensor are represented by heterotic supergravity fields as
\be\ba{rl}
S := &P^{\hM\hN} P^{\hP\hQ} S_{\hM\hP\hN\hQ} 
\\
	=& R - \tfrac{1}{12} H^2 + 4 \Box\phi -4 \partial_{\mu}\phi \partial^{\mu} \phi - \tfrac{1}{4}\alphap \Big( \tr F^2 - R^{(-)}{}_{\mu\nu mn} R^{(-)}{}^{\mu\nu mn}  \Big)\,.
\,,
\ea\ee
and
\be\ba{ll}
S_{m\brn} = S^{(0)}{}_{m\brn} + \frac{1}{4} \alphap \big( \cF_{m\brp}\big)^{\bra} \big( \cF_{\brn}{}^{\brp}\big)_{\bra} \,,
\\
S_{m \bra} = - \tfrac{1}{4\sqrt{2}} \alpha' D^{\brp} \big(\cF_{m \brp}\big)_{\bra} \,.
\ea\ee
Here $(\cF_{\mu\nu})_{\bra}$ is a field strength for the gauge field $\cA_{\mu\bra}$ which is decomposed as
\begin{equation}
  \left\{\begin{array}{ll}
  	(\cF_{\mu\nu})_{i} = (F_{\mu\nu})_{i}
  	\\
  	(\cF_{\mu\nu})_{[\brm\brn]} = R^{H}{}_{\mu\nu\brm\brn}
  \end{array}\right.
\end{equation}
where $(F_{\mu\nu})_i$ is field strength of Yang-Mills gauge field and  $R^{H}{}_{\mu\nu\brm\brn}$ is the curvature two-form in Riemannian geometry with the torsion $H$ which is defined in (\ref{RH}).
And $D_{\mu}$ is the covariant derivative in Riemannian geometry with a torsion $H$ 
\be\ba{ll}
D_{\mu} = \partial_{\mu} + \omega^H{}_{\mu} + \bromega^H{}_{\mu} 
\ea\ee

\newpage

\end{document}